\documentclass[preprint,aps,amsmath,byrevtex,prd,titlepage,
nofootinbib
,11pt
]
{revtex4-2}

\usepackage{dcolumn}
\usepackage{amsmath}
\usepackage{amssymb}
\usepackage{bm}
\usepackage{hyperref}
\usepackage{graphicx}
\usepackage{slashed}

\newcommand{\beq}{\begin{equation}}
\newcommand{\eeq}{\end{equation}}
\newcommand{\eq}[1]{Eq.~(\ref{#1})}

\begin{document}

\title{Hydrogen Energy Levels from the Anomalous Energy-Momentum QED Trace}

\author {Michael I. Eides}
\email[Email address: ]{meides@g.uky.edu}
\affiliation{Department of Physics and Astronomy,
University of Kentucky, Lexington, KY 40506, USA}

\begin{abstract}
Energy levels of hydrogen are calculated as one-loop  matrix elements of the QED  energy-momentum tensor trace in the external field approximation. An explicit connection established between the one-loop trace  diagrams and the standard Lamb shift one-loop diagrams. Our calculations provide an argument against inclusion of the
anomalous trace contribution as a separate term in the decomposition
of the QED quantum field Hamiltonian and serve as 
an illustration how the trace anomaly is realized in the bound state QED.

\end{abstract}

\maketitle


\section{Introduction}

Energy-momentum tensor (EMT) $T^{\mu\nu}$ describes interaction of fundamental particles and bound states with weak external gravitational field and was first discussed long time ago \cite{Kobzarev:1962wt,Pagels:1966zza}. Hadron EMT attracted a lot of attention and became an active field of experimental and theoretical research after it was discovered that, due to their connection with the generalized parton distribution functions, EMT form factors can be measured in deeply virtual Compton scattering and other hard exclusive reactions, see, e.g., \cite{Ji:1996ek,Ji:1996nm,Radyushkin:1996ru,Collins:1996fb,Kharzeev:1998bz,Berger:2001xd}. Low-energy QCD is nonperturbative, so, by necessity, nonperturbative methods and models are used in theoretical research on hadron EMT, see \cite{Ji:1994av,Ji:1995sv,Hudson:2017xug,Polyakov:2018zvc,Kharzeev:2021qkd,Liu:2021gco,
Lorce:2021xku,Ji:2020bby,Ji:2021pys,Ji:2021mtz,Metz:2020vxd} and references therein.

A new perspective on the EMT and its form factors could arise from consideration of the fundamental and bound states in quantum electrodynamics, where perturbative calculations are reliable. One can hope that comparison of the perturbative QED EMT with nonperturbative QCD EMT would lead to a deeper insight in both theories and the EMT properties. Perturbative EMT calculations were initiated in \cite{Milton:1971xnd,Milton:1973zz,Berends:1975ah,Milton:1976jr} and were further developed in recent papers \cite{Ji:1998bf,Kubis:1999db,Donoghue:2001qc, Rodini:2020pis,Sun:2020ksc,Ji:2021qgo,Metz:2021lqv,Ji:2021mfb,Ji:2022exr,Freese:2022jlu,
Eides:2023uox,Czarnecki:2023yqd,Czarnecki:2023dcv}, where a number of one-loop corrections to form factors, matrix elements and EMT trace for a free and bound electron were calculated.

We will discuss perturbative calculations of bound-state EMT trace below. It is well known that mass (rest energy) of any particle can be calculated as a diagonal matrix element of the EMT trace ${T^\mu}_\mu$ at rest, see, e.g., \cite{Ji:2021qgo,Dudas:1991kj} and references therein. Really, Hamiltonian is a three dimensional integral of  $T^{00}(x)$, $H=\int d^3xT^{00}(x)$, and then

\beq
\int d^3x\langle\bm p|T^{00}(x)|\bm p\rangle=E_{\bm p}\langle\bm p|\bm p\rangle,
\eeq

\noindent
where $|\bm p\rangle$ is a state with momentum $\bm p$ and $E_p$ is the the respective energy.

Due to translational invariance $\langle\bm p|T^{\mu\nu}(x)|\bm p\rangle=\langle\bm p|T^{\mu\nu}(0)|\bm p\rangle$, and hence

\beq
\langle\bm p|T^{00}(0)|\bm p\rangle=E_p\frac{\langle\bm p|\bm p\rangle}{V},
\eeq

\noindent
where $V$ is the space volume.

In covariant normalization $\langle\bm p|\bm p\rangle=2E_pV$  and $\langle\bm p|T^{00}(0)|\bm p\rangle=2E_p^2$. Due to Lorentz invariance $\langle\bm p|T^{\mu\nu}(0)|\bm p\rangle=2p^\mu p^\nu$ and the relationship $\int d^3x\langle\bm p|{T^\mu}_\mu(x)|\bm p\rangle=2m^2V$ holds for the EMT trace. In the rest frame $\int d^3x\langle\bm 0|{T^\mu}_\mu(x)|\bm 0\rangle=m\langle\bm 0|\bm 0\rangle$, and a normalization independent expression for the energy of any particle or system of particles with zero  total momentum has the form \cite{Dudas:1991kj}

\beq \label{traceqmass}
E =\frac{\int d^3x\langle\bm 0|{T^\mu}_\mu(x)|\bm 0\rangle}{\langle\bm 0|\bm 0\rangle}.
\eeq

\noindent
This is a universal formula valid in any quantum field theory both perturbatively and nonperturbatively.  We will use it in perturbation theory with nonrelativistic normalization.

Let us recall basics on EMT in gauge theories. EMT is a conserved operator and  $T_0^{\mu\nu}$ written in terms of bare fields (from which the bare or total Lagrangian is constructed) coincides with the renormalized EMT $[T^{\mu\nu}]_R$\footnote{We label renormalized local composite operators by the subscript $R$ below.}, which generates renormalized (UV finite) Green functions with renormalized fields $\phi_r$.
Due to the scale anomaly trace of EMT is nonzero even in QED and QCD with massless electrons (quarks) \cite{Hatta:2018sqd,Nielsen:1977sy,Adler:1976zt,Collins:1976yq,Minkowski:1976en,Tarrach:1981bi,Espriu:1982bw,
Freedman:1974gs,Freedman:1974ze}. In a massive theory

\beq \label{anomtrac}
{T_0^\mu}_\mu=[{T^\mu}_\mu]_R=(1+\gamma_m)[\bar\psi m\psi]_R+\frac{\beta(g)}{2g}[F^2]_R,
\eeq

\noindent
where $m$ is the mass of the fundamental fermion field in a theory under consideration, not the mass of a particle or bound state discussed above. The only difference between traces in an abelian and nonabelian (QED and QCD) theories is in the set of fermion fields and the form of the gauge field strengths. The left hand side of the trace equation  is renorminvariant and then the sum of the operators on the right hand side is also renorminvariant. The operator $\bar\psi_0 m_0\psi_0=m[\bar\psi \psi]_R$  is renorminvariant as a vertex in the Lagrangian. The sum of the remaining terms on the RHS, $\gamma_mm[\bar\psi\psi]_R+(\beta(g)/(2g)[F^2]_R$, is also renorminvariant,  see, e.g., \cite{Tarrach:1981bi}.


One can use perturbation theory and the explicit expression for the EMT trace in \eq{anomtrac} to calculate the matrix element \eq{traceqmass} in QED. The diagrams for the matrix element in \eq{traceqmass} do not coincide with the diagrams, which arise in perturbative calculations  of the same rest energy (mass) by more standard methods. While the anomaly theorem guarantees that both sets of diagrams lead to the same results, it could be interesting to check this coincidence by direct calculations and figure out which features of the two different sets of diagrams are responsible for this. We implemented this program in \cite{Eides:2023uox}, where we applied  \eq{traceqmass} to the one-loop mass renormalization of a free electron. We have calculated the sum of one-loop diagrams for the matrix element in \eq{traceqmass} and have shown that  the standard one-loop mass renormalization is reproduced in this way. We have also obtained an explicit analytic and diagrammatic relationships between two sets of diagrams, which explain why their sums are equal.

Below we will calculate energy levels of an electron bound in an external Coulomb field (hydrogen in the external field approximation) as matrix elements of the QED EMT trace in \eq{traceqmass}. We will use QED in the Furry picture \cite{Furry:1951bef,Sapirstein:1990gn,Weinberg:1995mt} and demonstrate how the Dirac-Coulomb energy levels  together with the  one-loop corrections  arise as matrix elements of the EMT trace. These one-loop corrections are just the well known contributions of order $\alpha(Z\alpha)^4m$  to the Lamb shift (to make the origin of the corrections more transparent we assume that the nucleus charge is $Ze$). Our goal is to trace out how and why two different sets of one-loop Feynman diagrams, one which arises in the classical Lamb shift calculations, and another, which contributes to the matrix element of the anomalous EMT trace,  produce coinciding results. In conclusion we will summarize the obtained results, compare them with the results of other authors and discuss further perspectives.

\section{Lamb shift in the Furry picture. Standard consideration}

\subsection{Furry picture}

The Furry picture \cite{Furry:1951bef,Sapirstein:1990gn,Weinberg:1995mt} is the most convenient framework for the discussion below. QED in the Furry picture is  quantized in the external Coulomb field, so the free electron field is expanded not in the plane waves, but in the eigenstates of the Dirac Hamiltonian in the external Coulomb field. One can use the ordinary Feynman diagram technique in the Furry picture, the only difference is that instead of the free electron propagator we should use the Dirac-Coulomb Green function

\beq
G=\frac{i}{p_0-\bm\alpha\cdot \bm p-\beta (m-i\epsilon)-V}\gamma_0
=\frac{i}{E-H}\gamma_0,
\eeq

\noindent
where $V=-Z\alpha/r$ is the Coulomb potential and $i$ in the numerator is included for consistency with the free Feynman propagator.

In terms of eigenfunctions the propagator has the form

\beq \label{dircoulgrfun}
G(E, \bm r,\bm r')
=\left\langle\bm r\left|\frac{i}{E-H}\gamma_0\right|\bm r'\right\rangle
=i\left[\sum_{n} \frac{\psi^{(+)}_n(\bm r) \bar  \psi^{(+)}_n(\bm r')}{E-E_n+i\epsilon}+\sum_{n} \frac{ \psi^{(-)}_n(\bm r) \bar\psi^ {(-)}_n(\bm r')}{E+E_n-i\epsilon}\right],
\eeq

\noindent
where summation goes over all states of discrete and continuous spectrum, $\psi^{(+)}_n(\bm r)$ and $\psi^{(-)}_n(\bm r)$ are eigenfunctions of the Dirac Hamiltonian in the external Coulomb field with positive and negative energies, respectively. These eigenfunctions are  normalized to one with the integration measure $\int d^3r$.

In this normalization \eq{traceqmass}  in the Furry picture turns into

\beq \label{basicmatrel}
E_n=\int d^3 r\langle n|{T^\mu}_\mu(0,\bm r)|n\rangle,
\eeq

\noindent
where $E_n$ is the energy of the electron in the bound state characterized by the multiindex $|n\rangle$ and normalized by the condition $\langle n'|n\rangle=\delta_{n'n}$.

\begin{figure}[h!]
\begin{center}
\includegraphics[height=3 cm]
{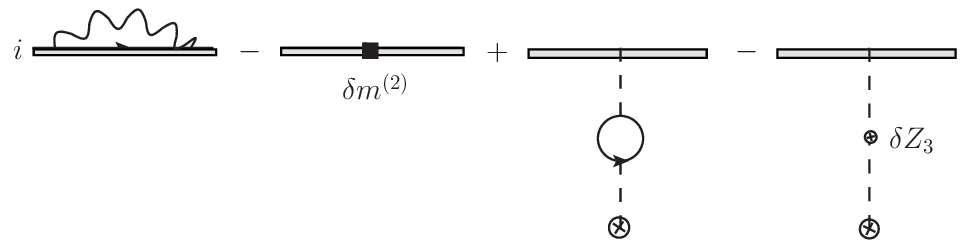}
\end{center}
\caption{Classical Lamb shift diagrams.}
\label{lshdfp}
\end{figure}

We start with the standard approach to the Lamb shift. Only two diagrams (and two counterterm diagrams) in Fig.~\ref{lshdfp} contribute to the one-loop Lamb shift of order $\alpha(Z\alpha)^4m$ if $Z\alpha\ll1$ in the Furry picture. Notice that it is sufficient to use the free electron propagator in the polarization loop, account for the binding effects in this loop generates contributions of higher orders in $Z\alpha$.
\subsection{Self-Energy Diagrams for the Lamb Shift in the Furry Picture}

\begin{figure}[h!]
\begin{center}
\includegraphics[height=1.5cm]
{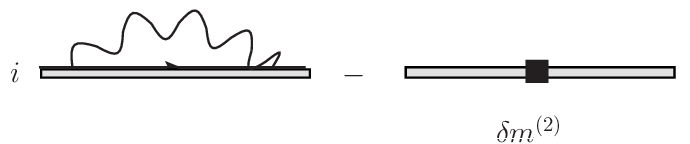}
\end{center}
\caption{Self-energy Lamb shift diagrams.}
\label{lshdfpse}
\end{figure}

The field theory matrix element of the leading self-energy (SE) contribution to the energy shift in Fig.~\ref{lshdfpse}  has the form (diagrammatically  $\Sigma=i\times$ $diagram$)

\beq \label{secontroenextf}
\Delta E_n^{SE}
=\int d^3 rd^3r'\langle n|\bar\psi(\bm r)\left[\Sigma_{reg}(\bm r,\bm r',E_n)-\delta^{(3)}(\bm r-\bm r') \delta m\right]\psi(\bm r')|n\rangle,
\eeq

\noindent
where $\Sigma_{reg}(\bm r,\bm r',E_n)$ is the regularized self-energy operator in the external Coulomb field, $\delta m=\Sigma^{(0)}_{reg}(\slashed p=m)$ and $\Sigma^{(0)}_{reg}(\slashed p=m)$ is the unrenormalized but regularized SE without external field\footnote{We use dimensional regularization and mass shell renormalization, respective formulae are collected in Appendix~\ref{renormsal}.}. The ultraviolet  (UV) divergences connected with the renormalization constant $Z_2$ are absent in the matrix element above, see, e.g., \cite{Sapirstein:1990gn}.

Calculating the QFT matrix elements above we obtain

\beq \label{selfenewimss}
\Delta E_n^{SE}
=\int d^3 rd^3r'\psi^\dagger_n(\bm r)\gamma^0\Sigma_{reg}(\bm r,\bm r',E_n)\psi_n(\bm r')-\delta m\int d^3r \psi^\dagger_n(\bm r)\gamma^0\psi_n(\bm r),
\eeq

\noindent
where $\psi_n(\bm r)$ are Dirac-Coulomb eigenfunctions normalized to one, $\int d^3r \psi^\dagger_n(\bm r)\psi_m(\bm r)=\delta_{nm}$.

After direct calculations the leading contribution to the Lamb shift due to the diagrams in Fig.~\ref{lshdfpse} \cite{{Sapirstein:1990gn},Weinberg:1995mt,blp1982}) can be represented as

\beq
\Delta E_{SE}(n,0)=\int d^3r\psi^\dagger_{n0}(\bm r)
\psi_{n0}(\bm r)V_{eff,se}(\bm r),
\eeq

\noindent
where

\beq \label{effselfenpo}
V_{eff,se}=\frac{4}{3}\frac{\alpha(Z\alpha)}{m^2}\left[\ln\frac{1}{(Z\alpha)^2}+\frac{5}{6}-\ln k_0(n,0)\right]\delta^{(3)}(\bm r),
\eeq

\noindent
where $\ln k_0(n,0)$ is the Bethe logarithm, $n$ is the principal quantum number and $\ell=0$ is the orbital momentum.

Respectively, the leading self-energy contribution to the Lamb shift is

\beq \label{onelplmbsgh}
\begin{split}
\Delta E_{SE}(n,0)&
=\frac{4}{3}\frac{\alpha(Z\alpha)}{m^2}\left[\ln\frac{1}{(Z\alpha)^2}+\frac{5}{6}-\ln k_0(n,0)\right]|\psi_{n0}(0)|^2\\
&=\frac{4}{3}\frac{\alpha(Z\alpha)^4m}{\pi n^3}\left[\ln\frac{1}{(Z\alpha)^2}+\frac{5}{6}-\ln k_0(n,0)
\right].
\end{split}
\eeq

\subsection{External Field Diagrams for the Lamb Shift}

\begin{figure}[h!]
\begin{center}
\includegraphics[height=2.5cm]
{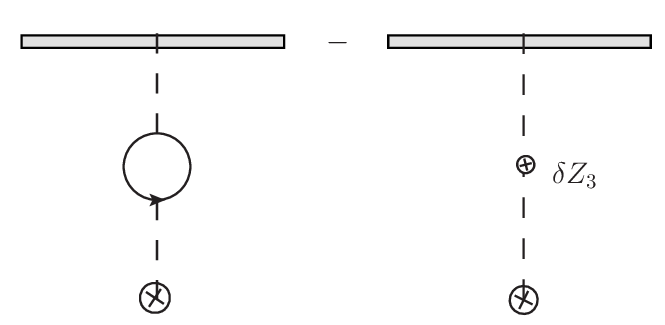}
\end{center}
\caption{External field Lamb shift diagrams.}
\label{lshdfpvp}
\end{figure}

The field Hamiltonian $H_{int}=\int d^3 x{\cal H}_{int}=e\int d^3 x\bar\psi\gamma_0\psi A^0_{ext}(x)$  describes interaction of the static external Coulomb field  with the electron. One-loop corrected static Coulomb field in Fig~\ref{lshdfpvp} has the form

\beq \label{extpotco}
A^0_{ext,one~loop}(\bm r)=-Ze\int\frac{d^3q}{(2\pi)^3}e^{i\bm q\cdot\bm r}\frac{\Pi_R(-\bm q^2)}{\bm q^2},
\eeq

\noindent
where  $\Pi_R(-\bm q^2)=\Pi_{reg}(-\bm q^2)-\Pi_{reg}(0)$, $\Pi_{reg}(0)=\delta Z_3$, see explicit expressions in Appendix~\ref{renormsal}.

The leading contribution to the Lamb shift arises from the low-$\bm q^2$ expansion of the renormalized polarization operator

\beq
\Pi^{(2)}_R(-\bm q^2)=-\frac{2\alpha}{\pi}\int_0^1dxx(1-x)\ln\frac{m^2}{x(1-x)\bm q^2+m^2}_{|\bm q^2/m^2\to0}
\to \frac{\alpha}{15\pi}\frac{\bm q^2}{m^2}.
\eeq

\noindent
Then the external field in \eq{extpotco} turns into

\beq \label{onelopexpolloq}
eA^0_{ext,one~loop}(\bm r)=-\frac{4\alpha(Z\alpha)}{15m^2}\delta^{(3)}(\bm r)\equiv V_{eff,pol}(\bm r),
\eeq

\noindent
and the leading external field (polarization loop) contribution to the Lamb shift is (see, e.g., \cite{{Sapirstein:1990gn},blp1982})

\beq \label{matrlelhintexf}
\begin{split}
\Delta& E_{VP}(n,\ell)=\langle n\ell|{H}_{int}|n\ell\rangle
=e\int d^3r\langle n\ell|\bar\psi(\bm r) \gamma_0\psi(\bm r)  A^0_{ext,one~loop}(\bm r)|n\ell\rangle\\
=&\int d^3r\psi^\dagger_{n\ell}(\bm r)
\psi_{n\ell}(\bm r)V_{eff,pol}(\bm r)
=-\frac{4\alpha(Z\alpha)}{15m^2}|\psi_{n\ell}(0)|^2
=-\frac{4\alpha(Z\alpha)^4m}{15\pi n^3}\delta_{\ell0}.
\end{split}
\eeq

\section{Hydrogen energy levels as  matrix elements of the EMT trace}

\subsection{EMT trace in one-loop approximation}

We are going to calculate one-loop matrix element of the EMT trace in \eq{basicmatrel}

\beq \label{exactre}
T=\int d^3r\langle e|m_0(1+\gamma_m(e_0))\bar\psi_0(\bm r)\psi_0(\bm r)+\frac{\beta(e_0)}{2e_0}F^2_0(\bm r)| e\rangle
\eeq

\noindent
in the Furry picture  for an electron in the external Coulomb field (hydrogen in the nonrecoil approximation).

\noindent
This matrix element in terms of renormalized fields in the one-loop approximation has the form

\beq \label{anomtrac2}
T\approx \int d^3r\langle e|[m-\delta m+m\gamma_m(e)+m\delta Z_2]\bar\psi(\bm r)\psi(\bm r)+\frac{\beta(e)}{2e}F^2(\bm r)| e\rangle.
\eeq

\subsubsection{Tree contribution}

In the leading approximation only the operator $m\bar\psi\psi$  in the trace in \eq{anomtrac2} gives contribution to the  matrix element in the hydrogen state.  Consider eigenstate $|{nj}\rangle$ which describes the Dirac-Coulomb  energy level $nj$. Then

\beq \label{treematreelm}
\int d^3r\langle {nj}|m\bar\psi(\bm r)\psi(\bm r)|{nj}\rangle=
m\int d^3r\psi_{nj}^\dagger(\bm r)\gamma^0\psi_{nj}(\bm r)=E_{nj},
\eeq

\noindent
where $E_{nj}$ is the exact eigenvalue of the Dirac Hamiltonian with the Coulomb external field

\beq \label{direignevalc}
E_{nj}=m\left[1+\left(\frac{Z\alpha}{n-\left(j+\frac{1}{2}\right)
+\sqrt{\left(j+\frac{1}{2}\right)^2-\left(Z\alpha\right)^2}}\right)^2\right]^{-\frac{1}{2}}.
\eeq

\noindent
The relationship in \eq{treematreelm} holds due to a relativistic virial theorem  derived by V.~A.~Fock at the dawn of Quantum Mechanics in 1930 \cite{fock1930}, for a later discussion, see, e.g., \cite{shabaev1991,shabaev2002}. We present a short derivation in Appendix~\ref{direnlev}.

\subsubsection{One-loop diagrams}

\begin{figure}[h!]
\begin{center}
\includegraphics[width=12cm]{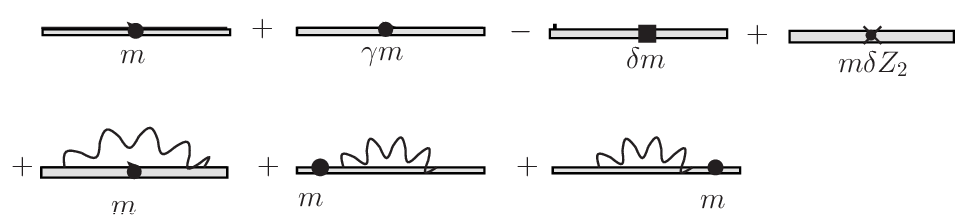}
\end{center}
\caption{Self-energy type trace Lamb shift diagrams.}
\label{hydemtdigse}
\end{figure}

\begin{figure}[h!]
\begin{center}
\includegraphics[height=2 cm]{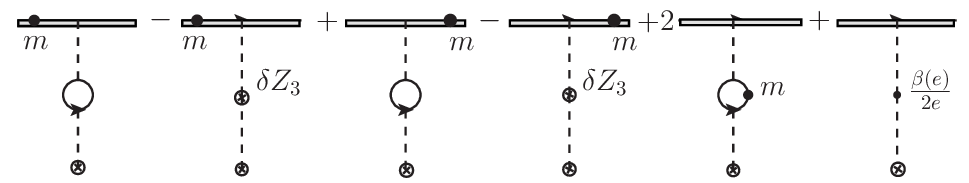}
\end{center}
\caption{Vacuum polarization type trace Lamb shift diagrams.}
\label{hydemtdigfp}
\end{figure}

Radiative corrections to the energy levels  corrections arise when we calculate matrix elements of the EMT trace in \eq{anomtrac2} beyond the tree approximation. Like in the standard calculation of the Lamb shift above  all diagrams for the matrix element of the  EMT trace in the one-loop approximation naturally split in two sets:  self-energy type diagrams  in Fig.~\ref{hydemtdigse}\footnote{We included in this set the tree diagram with the scalar vertex $m$, which generates the Dirac energy level. Notice also that the self-energy loops in this figure are subtracted, $\Sigma_{sub}=\Sigma_{reg}-\delta m$, because the counterterm contributions should be included.}  and vacuum polarization type diagrams Fig.~\ref{hydemtdigfp}.  Our first goal is to calculate all one-loop diagrams in Fig.~\ref{hydemtdigse} and Fig.~\ref{hydemtdigfp} and show that they reproduce the standard $\alpha(Z\alpha)^4m$ results in \eq{onelplmbsgh} and \eq{matrlelhintexf}.

The diagrams in Fig.~\ref{hydemtdigse} and Fig.~\ref{hydemtdigfp} do not coincide with the diagrams in Fig.~\ref{lshdfpse} and Fig.~\ref{lshdfpvp} and it is not obvious that they  produce the same results for the Lamb shift. Similar sets of different diagrams arise in  the case of a free electron in \cite{Eides:2023uox}, where the one-loop mass renormalization of a free electron was considered. The sum of the last three diagrams in the first row and the first diagram in the second row in Fig.~\ref{hydemtdigse} is zero for a free electron. The last two diagrams in the second row turn into zero on the mass shell for a free electron in the mass shell renormalization scheme. Therefore, the sum of all diagrams in Fig.~\ref{hydemtdigse} in the free case is $m$ on the mass shell, as it should be. There were no diagrams with an external field for a free electron.

We observed in \cite{Eides:2023uox} that in the free case logarithmic derivatives of the standard self-energy diagrams generate the diagrams for the trace. Due to linearity of the self-energy in mass both sets of diagrams lead to the same results. We expect that a similar mechanism will be at work for bound states.

We divided all one-loop diagrams for the matrix element of the EMT trace in \eq{anomtrac2} in two classes: diagrams with radiative insertions in the electron line (self-energy type diagrams) in Fig.~\ref{hydemtdigse}  and diagrams with radiative insertions in the external field (vacuum polarization type  diagrams) in Fig.~\ref{hydemtdigfp}. We will consider these gauge invariant  sets of diagrams separately.

\subsection{External field diagrams for the EMT trace }

\begin{figure}[h!]
\begin{center}
\includegraphics[height=2.5 cm]
{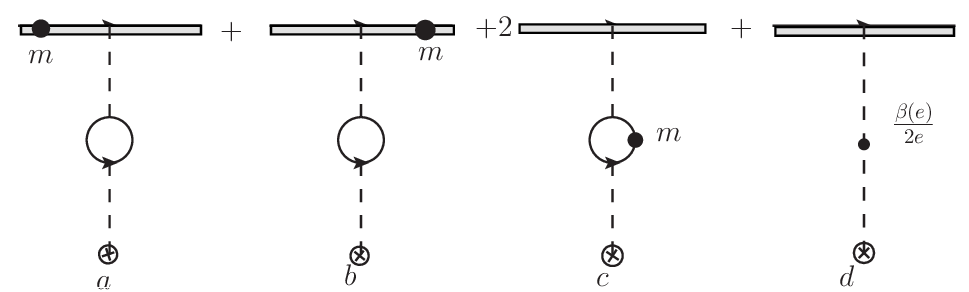}
\end{center}
\caption{External field trace Lamb shift diagrams.}
\label{vacpolmisel}
\end{figure}

Let us calculate six external field diagrams in Fig.~\ref{hydemtdigfp}. The first four diagrams arise as radiative corrections to the matrix element of  $m\bar\psi\psi$ in \eq{treematreelm}. Two diagrams with $\delta Z_3$ are due to the Lagrangian counterterm.  Then the first four diagrams in Fig.~\ref{hydemtdigfp} combine into two diagrams with the renormalized vacuum polarization operator, and we will use the standard one-loop expression for the renormalized polarization loop for their calculation. Thus we need to calculate the diagrams with external field in Fig.~\ref{vacpolmisel}: diagrams $(a)$ and $(b)$ with the renormalized one-loop insertion in the Coulomb photon and insertion of the scalar vertex $m\bar\psi\psi$ in the electron line, two diagrams $(c)$ with insertion of the scalar vertex $m\bar\psi\psi$ in the polarization loop and diagram $(d)$ with  $(\beta/2e)F^2$ insertion in the Coulomb photon.

\subsubsection{Matrix element of $m\bar\psi\psi$  with sidewise insertion of the polarization loop}

Diagrams $(a)$ and $(b)$ in Fig.~\ref{vacpolmisel} arise as one-loop perturbation theory corrections to the matrix element of the scalar vertex $m\bar\psi\psi$ in \eq{treematreelm}. Contributions of these diagrams to the energy shift are equal and  in the leading approximation can be written in the form

\beq \label{sidewisepolm}
\Delta E_a=\Delta E_b
=\int d^3rd^3r'\psi_{n}(\bm r)V_{eff,pol}(\bm r)[-iG_r(\bm r,\bm r',E_n)]m\gamma_0\psi_{n}(\bm r'),
\eeq

\noindent
where $G_r(E, \bm r,\bm r')$ is the reduced Dirac-Coulomb Green function (compare \eq{dircoulgrfun})

\beq \label{redgrefunc}
G_r(E, \bm r,\bm r')
=\left\langle\bm r\left|\left(\frac{i}{E-H}\right)'\gamma_0\right|\bm r'\right\rangle
=\left\langle\bm r\left|\sum_{k\neq n}\frac{i|k\rangle\langle k|}{E-E_k}\gamma_0\right|\bm r'\right\rangle,
\eeq

\noindent
and $V_{eff,pol}(\bm r)$ is defined in \eq{onelopexpolloq}.

The contributions in \eq{sidewisepolm} can be calculated  with the help of the virial relationships derived in \cite{shabaev1991,shabaev2002}. Respective calculations are rather cumbersome and we relegate their details to Appendix \ref{sideiwsepol}. After tedious calculations we obtain (see  \eq{sidwiswe})

\beq \label{sidewisepol}
\Delta E_a=\Delta E_b=-\frac{3}{2}\frac{4\alpha(Z\alpha)}{15 m^2}|\psi_{nl}(0)|^2=-\frac{3}{2}\frac{4\alpha(Z\alpha)^4m}{15 \pi n^3}\delta_{l0}
=\frac{3}{2}\Delta E_{VP}(n,\ell),
\eeq

\noindent
where $\Delta E_{VP}(n,\ell)$ is the total polarization contribution in \eq{matrlelhintexf}.

\subsubsection{Matrix element with the scalar vertex $m\bar \psi\psi$ insertion in the  polarization loop}

\begin{figure}[h!]
\begin{center}
\includegraphics[height=3cm]
{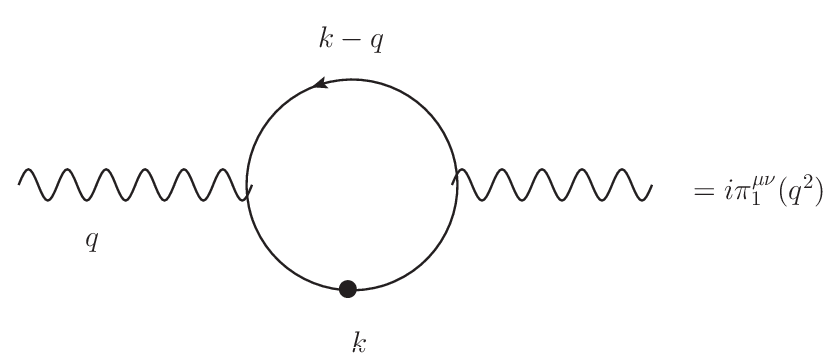}
\end{center}
\caption{Polarization loop with scalar vertex insertion.}
\label{polminsl}
\end{figure}

Contribution to the energy shift from the two identical diagrams $(c)$ in Fig.~\ref{vacpolmisel} has the form (nonrelativistic Schr\"odinger-Coulomb eigenfunctions are used below)

\beq  \label{masterfrompolm}
\Delta E_{c}=i4\pi Z\alpha \int d^3 r\psi^\dagger_{n\ell}(\bm r)\psi_{n\ell}(\bm r)
\int \frac{d^3q}{(2\pi)^3}e^{i\bm q\cdot\bm r}
\frac{2\pi_1(-\bm q^2)}{\bm q^2},
\eeq

\noindent
where the polarization loop with mass insertion $i\pi_1^{\mu\nu}(q)$ in Fig.~\ref{polminsl} is defined by the Feynman integral

\beq \label{ipi1polm}
i\pi_1^{\mu\nu}(q)=(-ie)^2(-1)m\int \frac{d^4k}{(2\pi)^4}Tr\left[\gamma^\mu
\left(\frac{i}{\slashed k-m+i\epsilon}\right)^2\gamma^\nu
\frac{i}{\slashed k-\slashed q-m+i\epsilon}\right].
\eeq

\noindent
Naively this integral is linearly divergent, but due to gauge invariance $i\pi_1^{\mu\nu}(q)=i(g^{\mu\nu}q^2-q^\mu q^\nu)\pi_1(q^2)$ and the remaining integral is convergent and does not require  any new counterterm. This is unlike the case of the standard polarization loop, where even after account for gauge invariance the logarithmic divergence survives and requires a counterterm. Calculating $\pi_1(q^2)$ we obtain

\beq \label{lowmomexppm}
\begin{split}
\pi_1(q^2)&=\frac{e^2m^2i}{\pi^2}\frac{1-\frac{4m^2 \tanh ^{-1}\left(\sqrt{\frac{-q^2}{4m^2-q^2}}\right)}{\sqrt{-q^2 (4 m^2-q^2)}}}{-2q^2}_{|q^2=-\bm q^2, \bm q^2/m^2\to0}\\
&\to \frac{2\alpha i}{\pi}\left(\frac{1}{6}-\frac{\bm q^2}{30m^2}+\frac{\bm q^4}{140m^4}+\ldots\right).
\end{split}
\eeq

\noindent
Next we plug this expansion in  \eq{masterfrompolm}, and arrive at the contribution to the energy level

\beq   \label{incompldic}
\Delta E_{c}
=-\frac{2\alpha(Z\alpha)^2m}{3\pi n^2}+\frac{8\alpha(Z\alpha)^4m}{15 \pi n^3}\delta_{\ell0}
=-\frac{2\alpha(Z\alpha)^2m}{3\pi n^2}-2\Delta E_{VP}(n,\ell).
\eeq

\noindent
The first term on the RHS arises after substitution  of the first term in the low-momentum expansion in \eq{lowmomexppm} in \eq{masterfrompolm}

\beq \label{nonerlcanterm}
-\frac{2\alpha(Z\alpha)}{3\pi}\int d^3 r\psi^\dagger_{nl}(\bm r)\frac{1}{r}\psi_{nl}(\bm r)=-\frac{2\alpha(Z\alpha)^2m}{3\pi n^2}.
\eeq

\noindent 
This term is of lower order in $Z\alpha$  than the leading contribution to the Lamb shift of order $\alpha(Z\alpha)^4m$, which we are calculating.  Hence, corrections of higher order in $Z\alpha$ to this term should be taken into account. We missed these corrections when we approximated Dirac-Coulomb wave functions by the  Schr\"odinger-Coulomb wave functions in \eq{masterfrompolm}(compare \cite{Sun:2020ksc,Ji:2021qgo}). Restoring the  Dirac-Coulomb wave functions we obtain  instead of \eq{nonerlcanterm} an exact in $Z\alpha$ result  (see, e.g.,\cite{shabaev2002})

\beq \label{exactinzaresm}
\begin{split}
\Delta E_{c1}&=-\frac{2\alpha(Z\alpha)}{3\pi}\int d^3 r\psi^\dagger_{njm}(\bm r)\frac{1}{r}\psi_{njm}(\bm r)\\
&=-\frac{2 \alpha (Z\alpha)^2 m }{3\pi}\frac{n-\left(j+\frac{1}{2}\right)+\frac{\left(j+\frac{1}{2}\right)^2}
{\sqrt{\left(j+\frac{1}{2}\right)^2-(Z\alpha)^2}}}{
   \left(\left[\sqrt{\left(j+\frac{1}{2}\right)^2-(Z\alpha)^2}+n-\left(j+\frac{1}{2}\right)
   \right]^2+(Z\alpha)^2\right)^{3/2}}\\
&\approx-\frac{2\alpha(Z\alpha)^2m}{3\pi n^2}+\frac{\alpha(Z\alpha)^4m}{\pi n^4}\left[1
-\frac{4n}{3\left(j+\frac{1}{2}\right)}\right]+\ldots.
\end{split}
\eeq

\noindent
We will discuss $\Delta E_{c1}$ below in connection with the anomaly term in Fig.~\ref{vacpolmisel} $(d)$.

Finally, the contribution to the energy shift from the two identical diagrams $(c)$ in Fig.~\ref{vacpolmisel} has the form 

\beq   \label{contrctt}
\Delta E_{c}
=\Delta E_{c1}-2\Delta E_{VP}(n,\ell).
\eeq

\noindent
Notice that the second term on the RHS in \eq{contrctt} is two times larger and has opposite sign to the total polarization contribution $\Delta E_{VP}(n,\ell)$ in \eq{matrlelhintexf}.

\subsubsection{Matrix element of the anomalous term $(\beta/2e)F^2$ insertion in the Coulomb photon}

Diagram $(d)$ in Fig.~\ref{vacpolmisel} arises as matrix element of the anomalous EMT term  $(\beta(e)/2e)F^2$ in \eq{anomtrac2}. This diagram  is similar to diagram $(c)$ in Fig.~\ref{vacpolmisel}, the only difference is that instead of insertion of the term $2i\pi_1^{\mu\nu}$ in the external Coulomb propagator in \eq{masterfrompolm}, we now insert the two-prong vertex $(\beta/2e)F^2$. In momentum space it has the form $4(\beta/(2e))(g_{\mu\nu}q^2-q_\mu q_\nu)$, which reduces to insertion of $4(\beta/(2e))$ in the Coulomb line. We use $4\beta(e)/2e=2\alpha/3\pi$ and Dirac-Coulomb wave functions to calculate the respective matrix element and obtain the result, which differs from the one in \eq{exactinzaresm} only by sign\footnote{This contribution was first calculated in eq.(7) in \cite{Sun:2020ksc} with two times larger  numerical factor and a wrong sign. The result above  agrees with the one in \cite{Ji:2021qgo}.}

\beq \label{anommamffelneq}
\Delta E_{d}=-\Delta E_{c1}.
\eeq

\noindent
Therefore, the contribution of the anomalous term in Fig.~\ref{vacpolmisel} $(d)$ exactly cancels with the one in \eq{exactinzaresm}. We will explain the reason for this cancellation below.

\subsubsection{Sum of all EMT trace polarization diagrams}

To calculate total contribution to the Lamb shift from the polarization type trace diagrams in Fig.~\ref{vacpolmisel} we collect contributions from the two diagrams with the sidewise insertions of the polarization perturbation to the scalar vertex in \eq{sidewisepol}, two  diagrams with the mass insertion in the polarization loop in \eq{contrctt}, and the diagram with the matrix element of the anomalous $(\beta/2e)F^2$ term in \eq{anommamffelneq}

\beq \label{sumofpoltrcom}
\Delta E=\Delta E_{a}+\Delta E_{b}+\Delta E_{c}+\Delta E_{d}=\Delta E_{VP}(n,\ell).
\eeq

\noindent
This is just the standard polarization contribution to the Lamb shift from \eq{matrlelhintexf}.

\subsection{EMT trace polarization type diagrams as derivatives of classical polarization diagrams}

\subsubsection{Heuristic considerations}

All contributions to the Lamb shift are linear in the electron mass $m$ and then $\Delta E_{SE}$ in \eq{onelplmbsgh} and $\Delta E_{VP}$ in \eq{matrlelhintexf} satisfy the relationship

\beq
\Delta E_n(m)=m\frac{d\Delta E_n(m)}{dm}.
\eeq

\noindent
So we expect (compare \cite{Eides:2023uox}) that the diagrams in Fig.~\ref{hydemtdigse} and Fig.~\ref{hydemtdigfp} originate as logarithmic mass derivatives $md/dm$ of the diagrams in Fig.~\ref{lshdfpse} and  Fig.~\ref{lshdfpvp}, respectively.

First we consider the one-loop polarization contribution to the Lamb shift in \eq{matrlelhintexf} and \eq{sumofpoltrcom}, which we calculated  from the diagrams in Fig.~\ref{lshdfpvp} and from a different set of diagrams in Fig.~\ref{vacpolmisel}. Of course, this is exactly what we had to expect from the trace anomaly, see \eq{traceqmass} and \eq{anomtrac}. Let us figure out analytically and diagrammatically what features of the two sets of diagrams are responsible for equality of their matrix elements. We return to \eq{matrlelhintexf}  and notice that the logarithmic mass derivative of

\beq \label{vacpolefplot}
\Delta E_{VP}(n,\ell)=-\frac{4}{15}\frac{\alpha(Z\alpha)}{m^2}|\psi_{n\ell}(0)|^2
\eeq

\noindent
can be written as

\beq \label{differevap}
\Delta E_{VP}(n,\ell)=
m\frac{\Delta E_{VP}(n,\ell)}{dm}
=2\frac{4}{15}\frac{\alpha(Z\alpha)}{m^2}|\psi_{n\ell}(0)|^2
-3\frac{4}{15}\frac{\alpha(Z\alpha)}{m^2}|\psi_{n\ell}(0)|^2,
\eeq

\noindent
where the two terms on the RHS arise from the $1/m^2$ prefactor and the wave function squared, respectively.  

The first term on the RHS in \eq{differevap} is equal to the sum of the diagrams $(c)$ and $(d)$ in Fig.~\ref{vacpolmisel} (see \eq{contrctt} and \eq{anommamffelneq}), and the second term is equal to the sum of the diagrams $(a)$ and $(b)$ in Fig.~\ref{vacpolmisel} with the sidewise mass insertions in the fermion line (see \eq{sidewisepol}). It remains to demonstrate that the diagrams in Fig.~\ref{vacpolmisel} arise as logarithmic mass derivatives of the diagrams in Fig.~\ref{lshdfpvp}. Calculating this derivative we need to remember about the bra and ket vectors in \eq{matrlelhintexf}, which are not shown explicitly in  Fig.~\ref{lshdfpvp}.

\subsubsection{Logarithmic derivative of the polarization loop in Fig.~\ref{lshdfpvp}\label{polarizader}}

The first term on the RHS in \eq{differevap} is a logarithmic derivative of the effective potential in \eq{matrlelhintexf} and we expect that it arises as the logarithmic derivative $md/dm$ of the polarization loop in Fig.~\ref{lshdfpvp}. Let us check this by direct calculation.  Logarithmic derivative of the polarization loop reduces to insertion of the scalar vertex $m$ in the propagators in the polarization loop and generates two identical diagrams $(c)$  in Fig.~\ref{vacpolmisel}. Notice that we differentiate regularized but not renormalized polarization operator. As we have seen considering the diagram in Fig.~\ref{polminsl}, this last diagram is UV convergent and does not require substraction. Hence, it should include a finite contribution from the finite logarithmic derivative of the logarithmically divergent polarization loop.

In dimensional regularization (see \eq{polareguidim})

\beq  \label{regpolopedofr}
\begin{split}
\Pi_{reg}^{(2)}(-\bm q^2)
&=-\frac{\alpha}{3\pi}\left[\frac{1}{\tilde\epsilon}+\ln\left(\frac{\mu^2}{m^2}\right)\right]
-\frac{2\alpha}{\pi}\int_0^1dxx(1-x)\ln\frac{m^2}{x(1-x)\bm q^2+m^2}\\
&
=\Pi_{reg}^{(2)}(0)+\Pi_{R}^{(2)}(-\bm q^2),
\end{split}
\eeq

\noindent
where in the mass shell renormalization scheme

\beq \label{conutertermpold}
\Pi_{reg}^{(2)}(0)=\delta Z_{3}=-\frac{\alpha}{3\pi}\left[\frac{1}{\tilde\epsilon}+\ln\left(\frac{\mu^2}{m^2}\right)\right],
\qquad \Pi_{R}^{(2)}(0)=0.
\eeq

\noindent
Respectively,  the logarithmic mass  derivative of the regularized polarization operator

\beq \label{logdiervapol}
\begin{split}
m\frac{d}{d m}\Pi_{reg}^{(2)}(-\bm q^2)
&\equiv m\frac{d}{d m}\Pi_{reg}^{(2)}(0)+m\frac{d}{d m}\Pi_{R}^{(2)}(-\bm q^2)\\
&
\to\frac{2\alpha}{3\pi}-\frac{2\alpha}{15\pi}\frac{\bm q^2}{m^2}.
\end{split}
\eeq

\noindent
Next we substitute  $\Pi_R(-\bm q^2)\to (m\partial/\partial m)\Pi_{reg}^{(2)}(-\bm q^2)$  in \eq{matrlelhintexf}  and obtain

\beq \label{dertermpollo1}
\begin{split}
\Delta E_{der}(n\ell)&\equiv
-4\pi(Z\alpha)\int d^3 r\psi^\dagger_{n\ell}(\bm r)\psi_{n\ell}(\bm r)\int\frac{d^3q}{(2\pi)^3}e^{i\bm q\cdot\bm r}\frac{m\frac{\partial\Pi_{reg}^{(2)}(-\bm q^2)}{\partial m}}{\bm q^2}\\
&
=-\frac{2\alpha(Z\alpha)^2m}{3\pi n^2}-2\Delta E_{VP}.
\end{split}
\eeq

\noindent
As expected, this result coincides with the result of the direct calculation of the two diagrams $(c)$ with mass insertions in the polarization loop in \eq{incompldic}. The first term on the RHS is due to ${d\Pi_{reg}^{(2)}(0)}/{d\ln m}$ in \eq{logdiervapol}. Like in the discussion after \eq{nonerlcanterm} in order to account for contributions of order $\alpha(Z\alpha)^4m$ and higher we need to restore the Dirac-Coulomb eigenfunctions in calculation of this term. Respective calculation reduces to the substitution of $\Delta E_{c1}$ from \eq{exactinzaresm} instead of the first term on the RHS in \eq{dertermpollo1}.

\subsubsection{Logarithmic derivative of the counterterm in Fig.~\ref{lshdfpvp}}

This time we apply the logarithmic derivative $md/dm$ to the counterterm $\delta Z_3$ in the second diagram in Fig.~\ref{lshdfpvp}

\beq \label{derpolrens}
m\frac{d\delta Z_3}{dm}=-\mu\frac{d\delta Z_3}{d\mu}
=\frac{2\beta(e)}{e}\approx \frac{2\alpha}{3\pi}.
\eeq

\noindent
The first equality on the RHS holds because the counterterm $\delta Z_3=\Pi_{reg}^{(2)}(0)$ is linear in $\ln(\mu/m)$, see \eq{conutertermpold}.

Now it is obvious that after differentiation of $\delta Z_3$ the second diagram in Fig.~\ref{lshdfpvp} turns into diagram $(d)$ in Fig.~\ref{vacpolmisel} with the matrix element of the anomalous term $(\beta/2e)F^2$ in the EMT trace and generates the $\Delta E_d$ contribution in \eq{anommamffelneq}.

The first term on the RHS in \eq{dertermpollo1} (and therefore in \eq{contrctt})   arises from the insertion of $m{d\delta Z_3}/{dm}$  in the external photon line. The respective diagram differs from the result of substitution $\delta Z_3\to d\delta Z_3/d\ln m$ in the second diagram  in Fig.~\ref{lshdfpvp} only by sign. We see that cancellation of $\Delta E_d$ contribution in \eq{anommamffelneq} and $\Delta E_{c1}$  in \eq{contrctt} (and in \eq{dertermpollo1}), which we observed above\footnote{This cancellation was also observed in \cite{Sun:2020ksc}.} is not accidental.  At the end of the day it is due to the definition of  the $\beta$-function.

\subsubsection{Logarithmic derivative of state vectors}

Contribution of the sum of the  diagrams  in Fig.~\ref{lshdfpvp} in \eq{matrlelhintexf}  has the form of a mass-dependent matrix element

\beq
\Delta E_n(m)=\langle n|Q(m)|n\rangle,
\eeq

\noindent
where matrix element $\Delta E_n(m)$ is a linear function of mass, and the Furry picture Dirac Hamiltonian eigenstates $|n\rangle$ and operator $Q(m)=H_{int}$ are some functions of the electron mass $m$. Then

\beq
\begin{split}
\Delta E_n(m)&=m\frac{d\Delta E_n(m)}{dm}\\
&
=\langle n|\left(m\frac{dQ(m)}{dm}\right)|n\rangle+
\left(m\frac{d}{dm}\langle n|\right)Q(m)|n\rangle+\langle n|Q(m)\left(m\frac{d }{dm}|n\rangle\right).
\end{split}
\eeq

\noindent
We have already calculated contribution of the first term on the RHS diagrammatically and analytically. It remains to consider the sum of two other terms. To calculate derivatives of state vectors we insert complete sets of states in the matrix elements

\beq
\begin{split}
&\left(m\frac{d}{dm}\langle n|\right)Q(m)|n\rangle+\langle n|Q(m)\left(m\frac{d}{dm}|n\rangle\right)
\\
&=\sum_k\left(m\frac{d}{dm}\langle n|\right)|k\rangle\langle k|Q(m)|n\rangle
+\sum_k\langle n|Q(m)|k\rangle\langle k|\left(m\frac{d}{dm}|n\rangle\right).
\end{split}
\eeq

\noindent
The term with $k=n$ does not contribute to the sums above

\beq
\begin{split}
&\left(m\frac{d}{dm}\langle n|\right)|n\rangle\langle n|Q(m)|n\rangle
+\langle n|Q(m)|n\rangle\langle n|\left(m\frac{d}{dm}|n\rangle\right)\\
&=\langle n|Q(m)|n\rangle\left[\left(m\frac{d}{dm}\langle n|\right)|n\rangle
+\langle n|\left(m\frac{d}{dm}|n\rangle\right)\right]\\
&
=\langle n|Q(m)|n\rangle m\frac{d\langle n|n\rangle}{dm}=0
\end{split}
\eeq

\noindent
since $\langle n|n\rangle=1$ is just the normalization condition. Then

\beq \label{twodirgpolmasin}
\begin{split}
&\left(m\frac{d}{dm}\langle n|\right)Q(m)|n\rangle+\langle n|Q(m)\left(m\frac{d}{dm}|n\rangle\right)\\
&=\sum_{k\neq n}\left(m\frac{d}{dm}\langle n|\right)|k\rangle\langle k|Q(m)|n\rangle
+\sum_{k\neq n}\langle n|Q(m)|k\rangle\langle k|\left(m\frac{d}{dm}|n\rangle\right).
\end{split}
\eeq

We use the Furry picture eigenvalue equation $H|n\rangle=E_n|n\rangle$, where $H$ is the Dirac Hamiltonian $H=\bm\alpha\cdot\bm p+\beta m+V$, to calculate the sums above. Matrix element $\langle k|H|n\rangle_{k\neq n}=0$ and hence at $k\neq n$

\beq
m\frac{d}{dm }\langle k|H|n\rangle=
E_n \left(m\frac{d}{dm}\langle k|\right)|n\rangle+\langle k|\beta m|n\rangle
+E_k\langle k|\left(m\frac{d}{dm} |n\rangle\right)=0,
\eeq

\noindent
and

\beq
m\frac{d}{dm}\langle k|n\rangle=\left(m\frac{d}{dm}\langle k|\right)|n\rangle+\langle k|\left(m\frac{d }{dm}|n\rangle\right)=0.
\eeq

\noindent
Combining these two equations we see that at $k\neq n$

\beq
\langle k|\left(m\frac{d }{dm}|n\rangle\right)=\frac{\langle k|\beta m|n\rangle}{E_n-E_k},
\eeq

\noindent
and

\beq
\sum_{k\neq n}|k\rangle\langle k|\left(m\frac{d}{dm}|n\rangle\right)=\sum_{k\neq n}|k\rangle\frac{\langle k|\beta m|n\rangle}{E_n-E_k}.
\eeq

\noindent
The reduced Green function in  \eq{redgrefunc} can be written as

\beq
G_r(E_n)=\left(\frac{i}{E_n-H}\right)'\gamma_0=\sum_{k\neq n}\frac{i|k\rangle \langle k|}{E_n-E_k}\gamma_0,
\eeq

\noindent
and the second sum on the RHS in \eq{twodirgpolmasin} has the form

\beq
\sum_{k\neq n}\langle n|Q(m)|k\rangle\langle k|\left(m\frac{d}{dm}|n\rangle\right)
=\langle n|Q(m)(-iG_r(E_n))m|n\rangle.
\eeq

\noindent
Comparing this expression  with \eq{sidewisepolm} we see that when $Q(m)=V_{eff,pol}$ this is exactly matrix element of the diagram $(b)$ in Fig.~\ref{vacpolmisel} with the sidewise insertion of the scalar vertex. Respectively, the first sum in \eq{twodirgpolmasin} describes diagram $(a)$ in Fig.~\ref{vacpolmisel}. We have already calculated matrix elements of these two diagrams in \eq{sidewisepol} (see also \eq{sidwiswe}) and observed that each of these diagrams contributes $(3/2)\Delta E_{VP}$.

Thus diagrams $(a)$ and $(b)$ in Fig.~\ref{vacpolmisel} arise as logarithmic derivatives of state vectors in the matrix elements of the Lamb shift polarization diagrams in Fig.~\ref{lshdfpvp}. Together with the results above this means that all polarization type trace diagrams in Fig.~\ref{vacpolmisel} arise as logarithmic derivatives of the Lamb shift polarization diagrams in Fig.~\ref{lshdfpvp}. This concludes our consideration of the trace diagrams in Fig.~\ref{vacpolmisel}. We have shown that their contribution to the energy of a bound state coincides with the contribution of the standard Lamb shift polarization diagrams in Fig.~\ref{lshdfpvp} and explained diagrammatically and analytically why contributions of these two different sets of diagrams coincide.

\subsection{EMT trace self-energy type diagrams as logarithmic derivatives of classical self-energy diagrams}

Tree and one-loop self-energy type diagrams for the EMT trace in Fig.~\ref{hydemtdigse} arise from the matrix element (compare  \eq{anomtrac2})

\beq
T\approx\int d^3r \langle e|[m-\delta m+m\gamma_m(e)+m\delta Z_2]\bar\psi(\bm r)\psi(\bm r)| e\rangle.
\eeq

\noindent
Similar diagrams were relevant for the discussion of a free electron mass renormalization in \cite{Eides:2023uox}.  The only modification is that the role of the propagator plays now the Dirac-Coulomb Green function in \eq{dircoulgrfun}. Notice that the $\delta m$  counterterm is included in the self-energy loops in the sidewise diagrams, which contain the subtracted Dirac-Coulomb  Green function.

\begin{figure}[h!]
\begin{center}
\includegraphics[width=8cm]{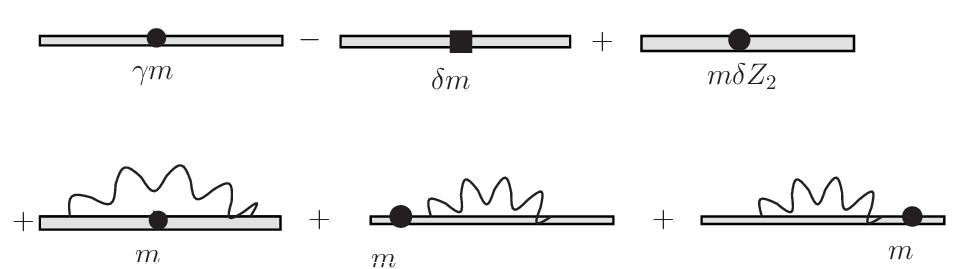}
\end{center}
\caption{One-loop self-energy trace diagrams.}
\label{hydemtdigse11}
\end{figure}

One-loop self-energy type trace diagrams in Fig.~\ref{lshdfpse} generate contributions to the Lamb shift. We are going to establish connection between the standard self-energy Lamb shift diagrams in Fig.~\ref{lshdfpse} and the self-energy type trace diagrams in Fig.~\ref{hydemtdigse11} and explain why they generate identical contributions.

\subsubsection{Cancellation of UV divergences}

\begin{figure}[h!]
\begin{center}
\includegraphics[width=10cm,height=1.5cm]{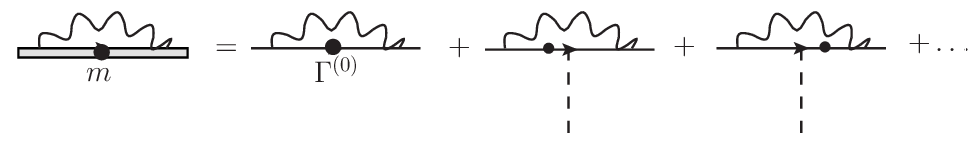}
\end{center}
\caption{External field expansion of one-loop scalar vertex.}
\label{hydemtdigse1}
\end{figure}

\begin{figure}[h!]
\begin{center}
\includegraphics[width=10cm,height=1.5cm]
{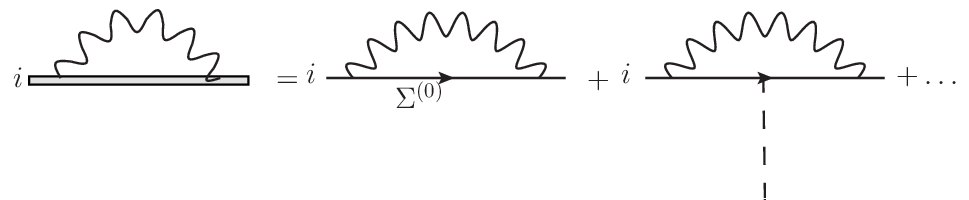}
\end{center}
\caption{External field expansion of one-loop self-energy.}
\label{Lambstandsexp}
\end{figure}

To use some results from \cite{Eides:2023uox} we consider external field expansions of the diagram with the scalar vertex  in Fig.~\ref{hydemtdigse1} and the self-energy diagram in Fig.~\ref{Lambstandsexp}. It was shown in  \cite{Eides:2023uox} that the one-loop self-energy $\Sigma^{(0)}(\slashed p)$ (without external field) and the respective mass renormalization term $\delta m^{(2)}=\Sigma^{(0)}(\slashed p=m)$ satisfy the relationships

\beq
m\frac{d\Sigma_{reg}^{(0)}(\slashed p=m)}{dm}=\Gamma^{(0)}(\slashed p=m)+m\delta Z_2,\quad
m\frac{d\delta m^{(2)}}{dm}=\delta m^{(2)}-m\gamma_m,
\eeq

\noindent
where $\Gamma^{(0)}_m(m)$ is the one-loop scalar vertex in Fig.~\ref{hydemtdigse1},  calculated with the free Feynman propagator instead of the Dirac-Coulomb propagator.

We see that the logarithmic mass derivative of  the leading term in the external field and on-mass-shell expansion of $(\Sigma_{reg}^{(0)}-\delta m^{(2)})$ in Fig.~\ref{lshdfpse} (the expression in the square brackets \eq{secontroenextf})

\beq \label{lagmasdivse}
m\frac{d\Sigma^{(0)}_{reg}(\slashed p=m)}{dm}-m\frac{d(\delta m^{(2)})}{dm}=\Gamma^{(0)}_m(m) +m\delta Z_2 -(\delta m^{(2)}-\gamma_mm)
\eeq

\noindent
generates the first three diagrams in Fig.~\ref{hydemtdigse11} and the leading term in the external field expansion of the fourth diagram calculated on shell.

For a bound electron we consider matrix elements of the diagrams Fig.~\ref{hydemtdigse11} between the Dirac-Coulomb eigenfunctions.  All effective vertices in \eq{lagmasdivse} are linear in the electron mass. We have seen in \eq{treematreelm} that the leading contribution of the scalar vertex $m$ to the energy level reduces to multiplication by the eigenvalue $E_{nj}$. Hence, the contribution to the energy level  from the first four diagrams in Fig.~\ref{hydemtdigse1} (with fourth diagram calculated on shell and without external field) is obtained from \eq{lagmasdivse} by multiplication by $E_{nj}$.

On the other hand $\delta m^{(2)}=\Sigma_{reg}^{(0)}(\slashed p=m)$ and the first three diagrams and the leading term in the external field expansion of the fourth diagram for the trace in Fig.~\ref{hydemtdigse11} cancel each other

\beq \label{cancellse}
-\delta m+m\gamma_m+m\delta Z_2=-\Gamma^{(0)}_m(m).
\eeq

\noindent
Hence, contribution to the energy shift from the first three diagrams in  Fig.~\ref{hydemtdigse1} plus the leading term in the external field expansion of the fourth diagram on shell is not only ultraviolet finite, but is equal zero\footnote{We can look at \eq{cancellse} from a slightly different perspective if we write it in the form

\beq \label{cancellserec}
\Gamma^{(0)}_m(m)+m\delta Z_2=\delta m-m\gamma_m.
\eeq

In this form it emphasizes that the last term on the RHS cancels  contribution of the first diagram in Fig.~\ref{hydemtdigse11}. We again observe cancellation of the anomalous contribution similar to the one in the case of the anomalous term $(\beta/2e)F^2$.}. The contribution of the remaining  diagrams in  Fig.~\ref{hydemtdigse11}  is ultraviolet  finite. As we will see below the contribution of the the fourth diagram in Fig.~\ref{hydemtdigse11} after subtraction of the first term in its external field expansion (see Fig.~\ref{hydemtdigse1}) is just $(-2\Delta  E_{SE})$.

\subsubsection{Calculation of the fourth diagram for the EMT trace in Fig.~\ref{hydemtdigse1}}

Expansion in the external field of the one-loop scalar vertex diagram in Fig.~\ref{hydemtdigse1} (fourth diagram in Fig.~\ref{hydemtdigse11}) starts with the diagram without external field. We just discussed that diagram and discovered that the UV divergent contribution cancels with the first three diagrams in Fig.~\ref{hydemtdigse1}.

Consider now next diagrams in the expansion of this diagram in the external field Fig.~\ref{hydemtdigse1}.  All these diagrams with one scalar vertex and any number of external field vertices arise from the first nontrivial diagram for the Lamb shift in Fig.~\ref{lshdfpse} after application  of the mass logarithmic derivative. Calculating this logarithmic derivative we differentiate the operator in \eq{secontroenextf} and not the state vectors. Using mass dependence of the explicit expression for the effective potential in \eq{effselfenpo} we see that  the subtracted fourth diagram in Fig.~\ref{hydemtdigse11} contributes $(-2\Delta  E_{SE})$ to the Lamb shift. This is the same mechanism as in the case of polarization operator, compare discussion in subsection \ref{polarizader} and \eq{onelopexpolloq}.

\subsubsection{Calculation of sidewise diagrams for the EMT trace in Fig.~\ref{hydemtdigse11}}

Like in the case of polarization operator sidewise diagrams in Fig.~\ref{hydemtdigse11} arise as derivatives of the state vectors and analytically the contribution to the Lamb shift of each of these diagrams has the form (compare \eq{sidewisepolm})

\beq
\Delta E=
\int d^3rd^3r'\psi_{n}(\bm r)V_{eff,se}(\bm r)[-iG_r(\bm r,\bm r',E_n)]m\gamma_0\psi_{n}(\bm r').
\eeq

\noindent
This contribution can be calculated analytically exactly like in Appendix \ref{sideiwsepol}. It is simpler to notice that the form of the polarization contribution to the Lamb shift in \eq{vacpolefplot} is similar to the self-energy contribution in \eq{onelplmbsgh}. Then we immediately conclude (compare \eq{differevap}) that each of the sidewise diagrams in  Fig.~\ref{hydemtdigse11} contributes $(3/2)\Delta  E_{SE}$ to the Lamb shift.

Finally, we have shown that the diagrams in Fig.~\ref{hydemtdigse11} arise as logarithmic derivatives of the self-energy Lamb shift diagrams in Fig.~\ref{lshdfpse} and the total leading order contribution of these diagrams, $-2\Delta E_{SE}+3\Delta E_{SE}=\Delta E_{SE}$, is the same.

\section{Summary}

We calculated the EMT trace contribution to the energy levels of hydrogen in the one-loop approximation. Graphically this contribution is represented by the diagrams in Fig.~\ref{hydemtdigse} and Fig.~\ref{hydemtdigfp}. The tree contribution of the scalar vertex $m$ is just the Dirac energy level in Coulomb field $E_{nj}$, see \eq{treematreelm}. The self-energy type trace diagrams in Fig.~\ref{hydemtdigse} and the polarization type trace diagrams in Fig.~\ref{hydemtdigfp} in the leading one-loop approximation  generate the well known self-energy and the polarization contributions to the Lamb shift  in \eq{onelplmbsgh} and \eq{matrlelhintexf}, respectively. In other words matrix element of the anomalous QED EMT trace in \eq{traceqmass} reproduces, as expected, hydrogen energy levels with account for the Lamb shift. Technically, the one-loop diagrams for the EMT trace arise as logarithmic mass derivatives of the standard Lamb shift diagrams in Fig.~\ref{lshdfp}. The only subtlety is that one needs to remember  to differentiate state vectors in the matrix elements. Equality of the contributions of the two sets of diagrams arises as a result of linearity in the electron mass of the hydrogen energy levels in the nonrecoil approximation.

Calculation of one-loop radiative corrections to the EMT trace for an electron in the Coulomb field was also addressed by other authors \cite{Sun:2020ksc,Ji:2021pys,Ji:2021qgo}. The derivative relationship between the diagrams in Fig.~\ref{lshdfp} and the EMT trace diagrams in Fig.~\ref{hydemtdigse} and Fig.~\ref{hydemtdigfp} was observed earlier from another perspective  in  \cite{Ji:2021qgo}. The  diagrams in Fig.~\ref{hydemtdigse} and Fig.~\ref{hydemtdigfp} with sidewise insertions of one-loop self-energy and polarization loop were missing in \cite{Sun:2020ksc}. The contributions of the diagrams calculated there were obtained with wrong coefficients and signs. As a result, the matrix element of the EMT trace calculated in \cite{Sun:2020ksc} did not reproduce  the classic expressions for the Lamb shift in \eq{onelplmbsgh} and \eq{matrlelhintexf}.

The matrix element of the anomalous part of the EMT trace $T_a=\int d^3r[\gamma_m m_0\bar\psi_0\psi_0+(\beta/2e_0)F^2_0]$ (sum of the second diagram in Fig.~\ref{hydemtdigse} and the last diagram in Fig.~\ref{hydemtdigfp}) was calculated earlier  in \cite{Ji:2021pys,Ji:2021qgo} is another way. In numerous works  $H_a=T_a/4$ was included as a separate term in the QED and QCD quantum field Hamiltonians and the respective contribution to the quantum state energy was called quantum anomalous energy, see, e.g.,\cite{Ji:2021mtz} and references therein. However, as we have seen above, dependence of the one-loop matrix elements of $H_a$  on the principal quantum number $n$ and the total electron angular momentum $j$ (see \eq{exactinzaresm}, the paragraph after \eq{lagmasdivse}, and also \cite{Ji:2021pys,Ji:2021qgo}) differs from the dependence of the one-loop Lamb shift contributions on these parameters.  Moreover, the sum of the second diagram in Fig.~\ref{hydemtdigse} and the last diagram in Fig.~\ref{hydemtdigfp} identically cancels with the contributions of other diagrams in these figures, and, hence, $H_a$  does not contribute to the one-loop shift of energy levels. Inclusion of $H_a$ in the decomposition of the quantum field Hamiltonian and, respectively, the proton mass, was a subject of active discussion in the literature, see, e.g., \cite{Ji:2021mtz,Ji:2021qgo,Hatta:2018sqd,Metz:2020vxd,Lorce:2021xku} and references therein. Wrong parametric dependence and complete cancellation of the quantum anomalous energy contribution to the hydrogen energy levels in the one-loop approximation indicate that decomposition of the QED Hamiltonian, which contains $H_a$  as a separate term is unwarranted. On the other hand, we expect that the QCD anomalous term $(\beta(g)/2g)F^2$  does not cancel and dominates in the chiral limit for light hadrons. Respectively, this could  justify decomposition of the  QCD Hamiltonian, which includes $H_a$, the anomalous part of the EMT trace, as a separate term. 

We believe that we presented above the first complete calculation  of the energy levels of hydrogen with account for one-loop corrections (Lamb shift) as matrix element of the EMT trace. We also explained diagrammatically and analytically why two different sets of perturbation theory  diagrams generate identical results.

There remains a number of open questions on the matrix element of the EMT trace as energy of a bound state at rest. It would be interesting to see how such matrix element reproduces hydrogen energy levels with account for hyperfine splitting and recoil, when there is a second mass parameter and as a result the energy levels are not linear in the electron mass any more.

With minor alterations the results above hold also for positronium. In both cases energy levels of a bound state are linear in the electron mass. The case of QCD is radically different. The chiral limit is a good approximation for the light hadrons, and the light quark masses give small contributions to their masses. The dominant contribution to the light hadrons masses  is determined by $\Lambda_{QCD}$. The EMT trace in QCD is similar to the one in QED (see \eq{anomtrac}) and its matrix element at rest is also equal to the mass of a bound state. Then we conclude that the dominant contribution to the light hadron masses is provided by the anomalous QCD EMT trace term $(\beta(g)/2g)F^2$, which cancelled in the  QED calculation above. On the other hand, in the case of a heavy quarkonium the dominant contribution to the quarkonium mass is supplied by the fermion contribution to the trace, proportional to the heavy quark mass. It would be interesting to trace out how relative weights of the fermion and gluon contributions to the mass of quarkonium change with decreasing quark mass, in other words the evolution from "bottomonium to $\rho$-meson". Of course, this cannot be done perturbatively, but the lattice gauge theory is probably an appropriate tool for this problem, see calculations in \cite{Yang:2014qna}. One could also try to make such calculations in QCD inspired strong interactions models, e.g., in the instanton liquid model \cite{Diakonov:1985eg}. We hope to address these open problems in the future.

\acknowledgments

I am grateful to Vladimir Yerokhin for explaining the formalism used in Appendix \ref{sideiwsepol}, to Andrzej Czarnecki and Keh-Fei Liu for useful discussions, and to Vladimir Pascalutsa for raising the problem of multiscale bound states in the context of the EMT trace. I thank Emilian  Dudas for informing me about \cite{Dudas:1991kj}, and Yizhuang Liu for a clarifying discussion on \cite{Ji:2021qgo}. This work was supported by the NSF grant  No. PHY-2011161.

\appendix

\section{One-loop renormalization constants\label{renormsal}}

We use dimensional regularization ($d=4-2\epsilon$) and mass-shell renormalization scheme. QED  Lagrangian in this scheme is

\beq
{\cal L}_0={\cal L}+\delta{\cal L}=-\frac{1}{4}F_0^2+\bar\psi_0(i\slashed\partial-m_0)\psi_0
-e_0\bar\psi_0\slashed A_0\psi_0,
\eeq

\noindent
where

\beq
\begin{split}
{\cal L}&=-\frac{1}{4}F^2+\bar\psi(i\slashed\partial-m)\psi
-\mu^\epsilon e\bar\psi\slashed A\psi, \\
%
%
\delta {\cal L}&=-\frac{1}{4}\delta Z_3F^2+\bar\psi(i\delta
Z_2\slashed\partial-\delta_m)\psi
-\mu^\epsilon e\delta Z_1\bar\psi\slashed A\psi.
\end{split}
\eeq

\noindent
The renormalization constants are defined as

\beq
\begin{split}
Z_1&=1+\delta Z_1,\quad Z_2=1+\delta Z_2, \quad Z_3=1+\delta Z_3, \quad
e_0=\mu^\epsilon Z_3^{-\frac{1}{2}}e,
\\
m_0&=mZ_mZ^{-1}_2,\quad
mZ_m=m(1+\delta Z_m)=m+\delta_m,\quad \delta m= m-m_0=m-mZ_mZ_2^{-1}.
\end{split}
\eeq

In the one-loop approximation

\beq \label{polareguidim}
\begin{split}
&\Pi_{reg}(q^2)=-\frac{2\alpha}{\pi}\int_0^1dxx(1-x)\left[\frac{1}{\tilde\epsilon}
+\ln\frac{\mu^2}{-x(1-x)q^2+m^2}
\right],\\
&\Sigma_{reg}(p)=\frac{\alpha}{2\pi}
\int_0^1dx\Biggl\{(2m-x\slashed p)
\left[\frac{1}{\tilde\epsilon}+
\ln\frac{\mu^2}{-x(1-x)p^2+x\lambda^2+(1-x)m^2}\right]
-(m-x\slashed p)\Biggr\},
\end{split}
\eeq

\noindent
where $\mu$ is the auxiliary dimensional regularization mass, $\lambda$ is the IR
photon mass, and $1/\tilde\epsilon={1}/{\epsilon}-\gamma+\ln(4\pi)$.

The one-loop counterterms  in the mass shell renormalization scheme are

\beq
\begin{split}
&\delta {Z_3}=\Pi_{reg}(0)=-\frac{\alpha}{3\pi}\left[\frac{1}{\tilde\epsilon}
+\ln\frac{\mu^2}{m^2}\right],\\
&m\delta Z_2-\delta_m=\Sigma_{reg}(m)
=\frac{3\alpha}{4\pi}m
\left[\frac{1}{\tilde\epsilon}+\ln\frac{\mu^2}{m^2}
+\frac{4}{3}\right]
\equiv \delta m,\\
&\delta Z_2=\Sigma'_{reg}(\slashed p=m)
=-\frac{\alpha}{4\pi}\left[\frac{1}{\tilde\epsilon}
+\ln\frac{\mu^2}{m^2}+2\ln\frac{\lambda^2}{m^2}+{4}\right],\\
&\delta Z_1=-\Lambda_{reg}(0)
=-\frac{\alpha}{4\pi}\left[\frac{1}{\tilde\epsilon}
+\ln\frac{\mu^2}{m^2}+2\ln\frac{\lambda^2}{m^2}+4\right],\\
&\delta_m\equiv\delta Z_m m=m\delta Z_2-\Sigma_{reg}(m)
=\frac{\alpha}{4\pi}m\left[-\frac{4 }{ \tilde\epsilon}- 2 \ln\left(\frac{\lambda^2}{m^2}\right)-4 \ln \left(\frac{\mu ^2}{m^2}\right)-8\right].
\end{split}
\eeq

\section{Relativistic Virial Theorem in Quantum Mechanics\label{direnlev}}

Let us prove that \cite{fock1930}

\beq \label{relvirth}
(nj|\beta m|nj)\equiv m\int d^3x\psi_{nj}^\dagger(x)\gamma^0\psi_{nj}(x)=E_{nj},
\eeq

\noindent
where we use quantum mechanical notation for  the state vectors, $|nj)$ is a Dirac-Coulomb eigenvector, $\psi_{nj}(x)$ and $E_{nj}$ are the respective eigenfunctions and eigenvalues.

This relationship follows from a relativistic virial theorem for the Dirac-Coulomb Hamiltonian
$H=\bm\alpha\cdot\bm p+\beta m+V(r)$, where $V(r)=-Z\alpha/r$.

We use the commutator relationships

\beq
[r_i,H]=[r_i, \bm\alpha\cdot\bm p]=i\alpha_i,\quad
[p_i,H]=[p_i,V(r)]=-i\partial_i V(r)_{|V=-Z\alpha/r}=-iZ\alpha\frac{r_i}{r^3},
\eeq

\noindent
to calculate the commutator $[\bm r\cdot\bm p,H]$.

Matrix element of this last commutator in an eigenstate of the Hamiltonian is zero, and we obtain

\beq
0=(nj|[\bm r\cdot\bm p,H]|nj)=(nj|\bm r\cdot[\bm p,H]|nj)+(nj|[\bm r,H]\cdot\bm p|nj)
\eeq
\[
=i\left[(nj| V(r)|nj)+(nj|\bm\alpha\cdot\bm p |nj)\right].
\]

\noindent
Then

\beq \label{forckres}
E_{nj}=(nj|H|nj)=(nj|\bm\alpha\cdot\bm p+\beta m+V|nj)=(nj|\beta m|nj),
\eeq

\noindent
Q.E.D.

In the nonrelativistic limit the relationship $\langle V(r)\rangle+\langle\bm\alpha\cdot\bm p \rangle=0$ reduces to the classical virial theorem $\langle V\rangle=-2\langle T\rangle$, compare \cite{Ji:2021qgo}.

\section{Calculation of the sidewise diagrams in Fig.~\ref{vacpolmisel} \label{sideiwsepol}}

We use virial relationships derived in \cite{shabaev1991,shabaev2002}  to calculate contribution of the diagrams with the sidewise insertion of the polarization leg. In the notation of \cite{shabaev1991,shabaev2002} (see also \cite{akhber1981,Mohr:1998grz,Rose:1961})  the eigenfunctions $\psi^\mu_{n\kappa}(\bm r)$ of the Dirac-Coulomb  equation

\beq
[\bm\alpha\cdot\bm p+\beta m+V(r)]\psi^\mu_{n\kappa}(\bm r)=E_{n\kappa}\psi^\mu_{n\kappa}(\bm r)
\eeq

\noindent
have the form

\beq
\psi^\mu_{n\kappa}(\bm r)=\left(
\begin{array}{c}
g_{n\kappa}(r)\chi_\kappa^\mu(\bm n)\\
if_{n\kappa}(r)\chi_{-\kappa}^\mu(\bm n)
\end{array}
\right),
\eeq

\noindent
where

\beq
\chi_\kappa^\mu(\bm n)=\sum_{m=\pm\frac{1}{2}}\left(\ell,\frac{1}{2},j|\mu-m,m\right)Y^{\mu-m}_\ell(\bm n)\chi^m,\qquad
\chi^\frac{1}{2}=
\left(
\begin{array}{c}
1\\
0
\end{array}\right),\qquad
\chi^{-\frac{1}{2}}=
\left(
\begin{array}{c}
0\\
1
\end{array}\right).
\eeq

\noindent
Here $\mu$ is the projection of the total angular momentum $j$, $\ell$ is the orbital momentum, $m=\pm 1/2$ is the projection of spin one half, integer $\kappa=(-1)^{j+\ell+1/2}(j+1/2)=\pm (j+1/2)\neq0$.  For $\kappa>0\Longrightarrow \ell=\kappa$ and for $\kappa<0\Longrightarrow \ell=-\kappa-1$.  Integer  $\kappa$ determines $\ell=|\kappa+1/2|-1/2$, and $j=|\kappa|-1/2$. In other words knowledge of $\kappa$ is equivalent to knowledge of $j$ and $\ell$, $\kappa\Leftrightarrow (j,\ell)$. The Clebsch-Gordan  coefficient $\left(\ell,\frac{1}{2},j|\mu-m,m\right)$ above in more standard notation is $\left(\ell,\mu-m;\frac{1}{2},m|j,\mu\right)$.

Shabaev \cite{shabaev1991} gets rid of angular dependence and works in terms of two-component "spinors" (radial functions) which turn into zero at $r=0$ and $r=\infty$

\beq
\phi_{n\kappa}(r)=\left(
\begin{array}{c}
 G_{n\kappa}(r)\\
 F_{n\kappa}(r)
 \end{array}
 \right),
\eeq

\noindent
where $G_{n\kappa}(r)=rg_{n\kappa}(r)$, $F_{n\kappa}(r)=rf_{n\kappa}(r)$. The functions $G_{n\kappa}(r)=rg_{n\kappa}(r)$ and  $F_{n\kappa}(r)=rf_{n\kappa}(r)$ are solutions of the system of two radial equations \cite{shabaev1991}

\beq
\begin{split}
&\frac{dG_{n\kappa}}{dr}+\frac{\kappa}{r}G_{n\kappa}-(E_{n\kappa}+m-V)F_{n\kappa}=0,\\
&\frac{dF_{n\kappa}}{dr}+\frac{\kappa}{r}F_{n\kappa}-(E_{n\kappa}-m-V)F_{n\kappa}=0.
\end{split}
\eeq

\noindent
The scalar product in the space of two-component functions $\phi_a(r)$ is defined as

\beq \label{shabscpr}
\langle \phi_a|\phi_b\rangle=\int_0^\infty dr(G_aG_b+F_aF_b).
\eeq

Following \cite{shabaev1991}  we will use below two-component radial states

\beq
|n\kappa\rangle=\left(
\begin{array}{c}
g_{n\kappa}(r)\\
f_{n\kappa}(r)
\end{array}
\right)
\eeq

\noindent
instead of four-component states $|n\ell m\rangle$.

We will also use special notation $|i,s,\kappa',n\kappa\rangle$ introduced in \cite{shabaev1991} for certain sums of matrix elements and states which resemble typical first order perturbation theory corrections
\beq \label{shabeaqa}
|i,s,\kappa',n\kappa\rangle=
\sum_{n'}^{E_{n'\kappa'}\neq E_{n\kappa}}\frac{|n'\kappa'\rangle\langle n'\kappa'|R_i^s|n\kappa\rangle}{E_{n\kappa}-E_{n'\kappa'}},
\eeq

\noindent
where $R_1^s=r^s$, $R_2^s=\sigma_z r^s$, $R_3^s=\sigma_xr^s$, $R_4^s=i\sigma_yr^s$.

The basic virial theorem in \eq{treematreelm} in this two-component notation has the form ($\gamma_0\to\sigma_3$)

\beq \label{direnetwocm}
\begin{split}
E_{nj}&=m\int d^3x\psi_{nj\ell}^\dagger(x)\gamma^0\psi_{nj\ell}(x)\\
&=m
\int d^3r
 \left(
\begin{array}{cc}
 g_{n\kappa}(r)\chi^{\mu\dagger}_{\kappa}(\bm n),&- if_{n\kappa}(r)\chi^{\mu\dagger}_{-\kappa}(\bm n)
 \end{array}
 \right)
 \left(
\begin{array}{cc}
 1&0\\
 0&-1
 \end{array}
 \right)
  \left(
\begin{array}{c}
 g_{n\kappa}(r)\chi_{\kappa}^\mu(\bm n)\\
 if_{n\kappa}(r)\chi_{-\kappa}^\mu(\bm n)
 \end{array}
 \right)\\
&=m\int_0^\infty dr r^2(g_{n\kappa}^2(r)-f_{n\kappa}^2(r))
=m\int_0^\infty dr (G_{n\kappa}^2(r)-F_{n\kappa}^2(r))\\
&=m\langle n\kappa|\sigma_3|n\kappa\rangle =mB^0_{n\kappa,n\kappa},
\end{split}
\eeq

\noindent
where (see  \cite{shabaev1991}) $B^0_{n\kappa,n\kappa}=E_{n\kappa}/m$.

To calculate the matrix element corresponding to one of the first two  diagrams in Fig.~\ref{vacpolmisel} we use perturbation theory expression in \eq{sidewisepolm}. The matrix element in this diagram is obviously symmetric with respect to the two-prong vertex $m\gamma_0$ and the perturbed Coulomb potential. We start considering $m\gamma_0$ as a perturbation. Then correction to the Dirac-Coulomb state vector has the from

\beq \label{pertdirnot}
|n\kappa\rangle^{(1)}=
\sum_{n'\neq n}\frac{|n'\kappa\rangle\langle n'\kappa|\gamma^0|n\kappa\rangle}{E_{n\kappa}-E_{n'\kappa}}.
\eeq

\noindent
Notice that (see \eq{direnetwocm})

\beq
\begin{split}
\langle n'jlm|\gamma_0|njlm\rangle&=
\int d^3 r
\left(
\begin{array}{cc}
 g_{n'\kappa}^\dagger(r)\chi^{\mu\dagger}_{\kappa}(\bm n)&
 -if_{n'\kappa}^\dagger(r)\chi^{\mu\dagger}_{-\kappa}(\bm n)
 \end{array}
 \right)
\left(
\begin{array}{cc}
 1&0\\
 0&-1
 \end{array}
 \right)
 \left(
\begin{array}{c}
 g_{n\kappa}(r)\chi^\mu_{\kappa}(\bm n)\\
 if_{n\kappa}(r)\chi^\mu_{-\kappa }(\bm n)
 \end{array}
 \right)\\
&=
\int_0^\infty r^2dr(g_{n'\kappa}g_{n\kappa}-f_{n'\kappa}f_{n\kappa})
=\langle n'\kappa|\sigma_3|n\kappa\rangle.
\end{split}
\eeq

\noindent
Then $|n\kappa\rangle^{(1)}$ in \eq{pertdirnot} in two-dimensional notation has the form

\beq
|n\kappa\rangle^{(1)}=\sum_{n'\neq n}\frac{|n'\kappa\rangle\langle n'\kappa|\sigma_3|n\kappa\rangle}{E_{n\kappa}-E_{n'\kappa}}
=|2,0,\kappa,n\kappa\rangle,
\eeq

\noindent
where at the last step we used \eq{shabeaqa}.

Next we use the expression for $|2,0,\kappa,n\kappa\rangle$ in Eq.(54) from \cite{shabaev2002}

\beq \label{shabaes2}
\begin{split}
|2,0,\kappa,n\kappa\rangle&=\frac{1}{m}\left(I-|n\kappa\rangle\langle n\kappa|\right)\left(E_{n\kappa}i\sigma_2 r+m\sigma_1 r +\alpha Zi\sigma_2-\kappa\sigma_3\right)|n\kappa\rangle\\
&=\frac{1}{m}\biggl[\left(E_{n\kappa}i\sigma_2 r+m\sigma_1 r +\alpha Zi\sigma_2-\kappa\sigma_3\right)|n\kappa\rangle\\
&-|n\kappa\rangle\langle n\kappa|\left(E_{n\kappa}i\sigma_2 r+m\sigma_1 r +\alpha Zi\sigma_2-\kappa\sigma_3\right)|n\kappa\rangle
\biggr].
\end{split}
\eeq

\noindent
Explicitly in the matrix form

\beq \label{matrnotshab}
\left(E_{n\kappa}i\sigma_2 r+m\sigma_1 r +\alpha Zi\sigma_2-\kappa\sigma_3\right)
= \left(
\begin{array}{cc}
 -\kappa &Er+mr+Z\alpha\\
 -Er+mr-Z\alpha&\kappa
 \end{array}
 \right).
\eeq

\noindent
The second term on the right hand side in \eq{shabaes2} is proportional to the expectation value

\beq
\langle n\kappa|\left(E_{n\kappa}i\sigma_2 r+m\sigma_1 r +\alpha Zi\sigma_2-\kappa\sigma_3\right)|n\kappa\rangle.
\eeq

\noindent
Expectation values of the terms proportional to $\sigma_2$ turn into zero

\beq
\langle n\kappa|(Er+Z\alpha)i\sigma_2|n\kappa\rangle=\int drr^2(Er+Z\alpha)(g_{n\kappa}f_{n\kappa}-g_{n\kappa}f_{n\kappa})=0,
\eeq

\noindent
and the contribution of the other terms in \eq{matrnotshab} is

\beq
\langle n\kappa|\left(m\sigma_1 r -\kappa\sigma_3\right)|n\kappa\rangle
=C^1_{n\kappa,n\kappa}-\kappa B^0_{n\kappa,n\kappa},
\eeq

\noindent
where \cite{shabaev2002}

\beq
C^s_{n\kappa,n\kappa}=2\int_0^\infty dr r^s G_{n\kappa}F_{n\kappa}, \qquad
B^s_{n\kappa,n\kappa}=\int_0^\infty dr r^s (G^2_{n\kappa}-F^2_{n\kappa}).
\eeq

\noindent
Explicitly (see \cite{shabaev2002}) $C^1_{n\kappa,n\kappa}=(2\kappa E_{n\kappa}-m)/2m^2$ and $B^0_{n\kappa,n\kappa}=E_{n\kappa}/m$, and then

\beq
\begin{split}
&\langle n\kappa|\left(E_{n\kappa}i\sigma_y r+m\sigma_x r +\alpha Zi\sigma_y-\kappa\sigma_z\right)|n\kappa\rangle\\
&=C^1-\kappa B^0=B^0=\frac{2\kappa E_{n\kappa}-m}{2m^2}-\frac{\kappa E_{n\kappa}}{m}=-\frac{1}{2m}.
\end{split}
\eeq

\noindent
Then $|2,0,\kappa,n\kappa\rangle$ in \eq{shabaes2} simplifies and in the matrix notation with account for \eq{matrnotshab} has the form

\beq
\begin{split}
m|2,0,\kappa,n\kappa\rangle&=
\left(E_{n\kappa}i\sigma_y r+m\sigma_x r +\alpha Zi\sigma_y-\kappa\sigma_z\right)|n\kappa\rangle
+\frac{1}{2}|n\kappa\rangle\\
&=
\left(
\begin{array}{cc}
 -\kappa+\frac{1}{2} &E_{n\kappa}r+mr+Z\alpha\\
 -E_{n\kappa}r+mr-Z\alpha&\kappa+\frac{1}{2}
 \end{array}
 \right)
 |n\kappa\rangle\\
&=
\left(
\begin{array}{cc}
 -\kappa+\frac{1}{2} &(E_{n\kappa}+m)r+Z\alpha\\
 -(E_{n\kappa}-m)r-Z\alpha&\kappa+\frac{1}{2}
 \end{array}
 \right)|n\kappa\rangle
 =
 \left(
\begin{array}{c}
 \tilde g_{n\kappa}\\
 \tilde f_{n\kappa}
 \end{array}
 \right),
 \end{split}
\eeq

\noindent
where

\beq
\left(
\begin{array}{c}
 \tilde g_{n\kappa}\\
 \tilde f_{n\kappa}
 \end{array}
 \right)
 =
 \left(
\begin{array}{c}
 (-\kappa+\frac{1}{2})g_{n\kappa}+ [(E_{n\kappa}+m)r+Z\alpha]f_{n\kappa}\\
 (\kappa+\frac{1}{2})f_{n\kappa} -[(E_{n\kappa}-m)r-Z\alpha]g_{n\kappa}
 \end{array}
 \right).
\eeq

\noindent
In the four-component notation $m|2,0,\kappa,n\kappa\rangle$ the last two-component state has the form

\beq
 \left(
\begin{array}{c}
 \tilde g_{n\kappa} \chi^\mu_\kappa\\
 \tilde f_{n\kappa}\chi^\mu_{-\kappa}
 \end{array}
 \right).
\eeq

\noindent
We are calculating matrix element

\beq
\Delta E_{m,pol}=\langle n\kappa|V_{eff,pol}m|2,0,\kappa,n\kappa\rangle,
\eeq

\noindent
where the radiatively corrected Coulomb potential  in the nonrelativistic approximation has the form (see \eq{onelopexpolloq})

\beq \label{polapotvs}
V_{eff,pol}(r)=-\frac{4\alpha(Z\alpha)}{15 m^2}\delta^{(3)}(\bm r).
\eeq

\noindent
In the nonrelativistic approximation and in the leading order in $Z\alpha$  we preserve only the large component $g_{n\kappa}$ which turns into the Schrodinger-Coulomb wave function $g_{n\kappa}\Omega_{\kappa m}\to\psi_{nlm}$ and calculating the matrix element obtain

\beq
\Delta E_{m,pol}=-\left(-\kappa+\frac{1}{2}\right)\frac{4\alpha(Z\alpha)}{15 m^2}|\psi_{nl}(0)|^2.
\eeq

\noindent
This contribution is nonzero only for  $s$-states and $\kappa=-1$ for all $s$-states. Finally, we obtain contributions of each of the first two diagrams in Fig.~\ref{vacpolmisel}

\beq \label{sidwiswe}
\Delta E_a=\Delta E_b=-\frac{3}{2}\frac{4\alpha(Z\alpha)^4m}{15 \pi n^3}\delta_{l0}
=\frac{3}{2}\Delta E_{VP}(n,\ell),
\eeq

\noindent
where $\Delta E_{VP}(n,\ell)$ is the total polarization contribution in \eq{matrlelhintexf}.


\begin{thebibliography}{99}

\bibitem{Kobzarev:1962wt}
I.~Y.~Kobzarev and L.~B.~Okun,
``Gravitational interaction of fermions,''
Zh. Eksp. Teor. Fiz. \textbf{43}, 1904-1909 (1962).


\bibitem{Pagels:1966zza}
H.~Pagels,
''Energy-Momentum Structure Form Factors of Particles,''
Phys. Rev. \textbf{144}, 1250-1260 (1966)
doi:10.1103/PhysRev.144.1250.

\bibitem{Ji:1996ek}
X.~D.~Ji,
''Gauge-Invariant Decomposition of Nucleon Spin,''
Phys. Rev. Lett. \textbf{78}, 610-613 (1997)
doi:10.1103/PhysRevLett.78.610
[arXiv:hep-ph/9603249 [hep-ph]].


\bibitem{Ji:1996nm}
X.~D.~Ji,
''Deeply virtual Compton scattering,''
Phys. Rev. D \textbf{55}, 7114-7125 (1997)
doi:10.1103/PhysRevD.55.7114
[arXiv:hep-ph/9609381 [hep-ph]].

\bibitem{Radyushkin:1996ru}
A.~V.~Radyushkin,
''Asymmetric gluon distributions and hard diffractive electroproduction,''
Phys. Lett. B \textbf{385}, 333-342 (1996)
doi:10.1016/0370-2693(96)00844-1
[arXiv:hep-ph/9605431 [hep-ph]].

\bibitem{Collins:1996fb}
J.~C.~Collins, L.~Frankfurt and M.~Strikman,
``Factorization for hard exclusive electroproduction of mesons in QCD,''
Phys. Rev. D \textbf{56}, 2982-3006 (1997)
doi:10.1103/PhysRevD.56.2982
[arXiv:hep-ph/9611433 [hep-ph]].

\bibitem{Kharzeev:1998bz}
D.~Kharzeev, H.~Satz, A.~Syamtomov and G.~Zinovjev,
''$J/\psi$ photoproduction and the gluon structure of the nucleon,''
Eur. Phys. J. C \textbf{9}, 459-462 (1999)
doi:10.1007/s100529900047
[arXiv:hep-ph/9901375 [hep-ph]].

\bibitem{Berger:2001xd}
E.~R.~Berger, M.~Diehl and B.~Pire,
''Time-like Compton scattering: Exclusive photoproduction of lepton pairs,''
Eur. Phys. J. C \textbf{23}, 675-689 (2002)
doi:10.1007/s100520200917
[arXiv:hep-ph/0110062 [hep-ph]].

\bibitem{Ji:1994av}
X.~D.~Ji,
``A QCD analysis of the mass structure of the nucleon,''
Phys. Rev. Lett. \textbf{74}, 1071-1074 (1995)
doi:10.1103/PhysRevLett.74.1071
[arXiv:hep-ph/9410274 [hep-ph]].


\bibitem{Ji:1995sv}
X.~D.~Ji,
``Breakup of hadron masses and energy - momentum tensor of QCD,''
Phys. Rev. D \textbf{52}, 271-281 (1995)
doi:10.1103/PhysRevD.52.271
[arXiv:hep-ph/9502213 [hep-ph]].

\bibitem{Hudson:2017xug}
J.~Hudson and P.~Schweitzer,
``D term and the structure of pointlike and composed spin-0 particles,''
Phys. Rev. D \textbf{96}, no.11, 114013 (2017)
doi:10.1103/PhysRevD.96.114013
[arXiv:1712.05316 [hep-ph]].

\bibitem{Polyakov:2018zvc}
M.~V.~Polyakov and P.~Schweitzer,
''Forces inside hadrons: pressure, surface tension, mechanical radius, and all that,''
Int. J. Mod. Phys. A \textbf{33}, no.26, 1830025 (2018)
doi:10.1142/S0217751X18300259
[arXiv:1805.06596 [hep-ph]].

\bibitem{Kharzeev:2021qkd}
D.~E.~Kharzeev,
``Mass radius of the proton,''
Phys. Rev. D \textbf{104}, no.5, 054015 (2021)
doi:10.1103/PhysRevD.104.054015
[arXiv:2102.00110 [hep-ph]].

\bibitem{Liu:2021gco}
K.~F.~Liu,
``Proton mass decomposition and hadron cosmological constant,''
Phys. Rev. D \textbf{104}, no.7, 076010 (2021)
doi:10.1103/PhysRevD.104.076010
[arXiv:2103.15768 [hep-ph]].

\bibitem{Lorce:2021xku}
C.~Lorc\'e, A.~Metz, B.~Pasquini and S.~Rodini,
``Energy-momentum tensor in QCD: nucleon mass decomposition and mechanical equilibrium,''
JHEP \textbf{11}, 121 (2021)
doi:10.1007/JHEP11(2021)121
[arXiv:2109.11785 [hep-ph]].

\bibitem{Ji:2020bby}
X.~Ji, Y.~Liu and I.~Zahed,
``Mass structure of hadrons and light-front sum rules in the $'t$ Hooft model,''
Phys. Rev. D \textbf{103}, no.7, 074002 (2021)
doi:10.1103/PhysRevD.103.074002
[arXiv:2010.06665 [hep-ph]].

\bibitem{Ji:2021pys}
X.~Ji and Y.~Liu,
``Quantum anomalous energy effects on the nucleon mass,''
Sci. China Phys. Mech. Astron. \textbf{64}, no.8, 281012 (2021)
doi:10.1007/s11433-021-1723-2
[arXiv:2101.04483 [hep-ph]].

\bibitem{Ji:2021mtz}
X.~Ji,
``Proton mass decomposition: naturalness and interpretations,''
Front. Phys. (Beijing) \textbf{16}, no.6, 64601 (2021)
doi:10.1007/s11467-021-1065-x
[arXiv:2102.07830 [hep-ph]].

\bibitem{Metz:2020vxd}
A.~Metz, B.~Pasquini and S.~Rodini,
``Revisiting the proton mass decomposition,
Phys. Rev. D \textbf{102}, 114042 (2020)
doi:10.1103/PhysRevD.102.114042
[arXiv:2006.11171 [hep-ph]].



\bibitem{Milton:1971xnd}
K.~A.~Milton,
``Quantum corrections to stress tensors and conformal invariance,''
Phys. Rev. D \textbf{4}, 3579-3593 (1971)
doi:10.1103/PhysRevD.4.3579.

\bibitem{Milton:1973zz}
K.~A.~Milton,
``Scale invariance and spectral forms for conformal stress tensors. (erratum),''
Phys. Rev. D \textbf{7}, 1120 (1973)
[erratum: Phys. Rev. D \textbf{7}, 3821 (1973)]
doi:10.1103/PhysRevD.7.1120.

\bibitem{Berends:1975ah}
F.~A.~Berends and R.~Gastmans,
``Quantum Electrodynamical Corrections to Graviton-Matter Vertices,''
Annals Phys. \textbf{98}, 225 (1976)
doi:10.1016/0003-4916(76)90245-1.

\bibitem{Milton:1976jr}
K.~A.~Milton,
``Quantum Electrodynamic Corrections to the Gravitational Interaction of the electron,''
Phys. Rev. D \textbf{15}, 538 (1977)
doi:10.1103/PhysRevD.15.538.


\bibitem{Ji:1998bf}
X.~D.~Ji and W.~Lu,
''A Modern anatomy of electron mass,''
[arXiv:hep-ph/9802437 [hep-ph]].

\bibitem{Kubis:1999db}
B.~Kubis and U.~G.~Meissner,
``Virtual photons in the pion form-factors and the energy momentum tensor,''
Nucl. Phys. A \textbf{671}, 332-356 (2000)
[erratum: Nucl. Phys. A \textbf{692}, 647-648 (2001)]
doi:10.1016/S0375-9474(99)00823-4
[arXiv:hep-ph/9908261 [hep-ph]].

\bibitem{Donoghue:2001qc}
J.~F.~Donoghue, B.~R.~Holstein, B.~Garbrecht and T.~Konstandin,
``Quantum corrections to the Reissner-Nordstr\"om and Kerr-Newman metrics,''
Phys. Lett. B \textbf{529}, 132-142 (2002)
[erratum: Phys. Lett. B \textbf{612}, 311-312 (2005)]
doi:10.1016/S0370-2693(02)01246-7
[arXiv:hep-th/0112237 [hep-th]].

\bibitem{Rodini:2020pis}
S.~Rodini, A.~Metz and B.~Pasquini,
''Mass sum rules of the electron in quantum electrodynamics,''
JHEP \textbf{09}, 067 (2020)
doi:10.1007/JHEP09(2020)067
[arXiv:2004.03704 [hep-ph]].

\bibitem{Sun:2020ksc}
B.~d.~Sun, Z.~h.~Sun and J.~Zhou,
``Trace anomaly contribution to hydrogen atom mass,''
Phys. Rev. D \textbf{104}, no.5, 056008 (2021)
doi:10.1103/PhysRevD.104.056008
[arXiv:2012.09443 [hep-ph]].

\bibitem{Ji:2021qgo}
X.~Ji, Y.~Liu and A.~Sch\"afer,
''Scale symmetry breaking, quantum anomalous energy and proton mass decomposition,''
Nucl. Phys. B \textbf{971}, 115537 (2021)
doi:10.1016/j.nuclphysb.2021.115537
[arXiv:2105.03974 [hep-ph]].




\bibitem{Metz:2021lqv}
A.~Metz, B.~Pasquini and S.~Rodini,
``The gravitational form factor D(t) of the electron,''
Phys. Lett. B \textbf{820}, 136501 (2021)
doi:10.1016/j.physletb.2021.136501
[arXiv:2104.04207 [hep-ph]].

\bibitem{Ji:2021mfb}
X.~Ji and Y.~Liu,
``Momentum-Current Gravitational Multipoles of Hadrons,''
Phys. Rev. D \textbf{106}, no.3, 034028 (2022)
doi:10.1103/PhysRevD.106.034028
[arXiv:2110.14781 [hep-ph]].

\bibitem{Ji:2022exr}
X.~Ji and Y.~Liu,
``Gravitational Tensor-Monopole Moment of Hydrogen Atom To Order $\mathcal{O}(\alpha)$,''
[arXiv:2208.05029 [hep-ph]].

\bibitem{Freese:2022jlu}
A.~Freese, A.~Metz, B.~Pasquini and S.~Rodini,
``The gravitational form factors of the electron in quantum electrodynamics,
Phys. Lett. B \textbf{839}, 137768 (2023)
doi:10.1016/j.physletb.2023.137768
[arXiv:2212.12197 [hep-ph]].


\bibitem{Eides:2023uox}
M.~I.~Eides,
``One-loop electron mass and QED trace anomaly,''
Eur. Phys. J. C \textbf{83}, no.5, 356 (2023)
doi:10.1140/epjc/s10052-023-11535-6
[arXiv:2301.02883 [hep-ph]].


\bibitem{Czarnecki:2023yqd}
A.~Czarnecki, Y.~Liu and S.~N.~Reza,
''Energy-momentum Tensor of a Hydrogen Atom: Stability, $D$-term, and the Lamb Shift,''
Acta Phys. Polon. Supp. \textbf{16}, no.7, 7-A19 (2023)
doi:10.5506/APhysPolBSupp.16.7-A19
[arXiv:2309.10994 [hep-ph]].

\bibitem{Czarnecki:2023dcv}
A.~Czarnecki,
``Radiative corrections to mechanical properties of bound states,''
PoS \textbf{RADCOR2023}, 094 (2024)
doi:10.22323/1.432.0094.

\bibitem{Dudas:1991kj}
E.~A.~Dudas and D.~Pirjol,
``A Virial theorem in quantum field theory,''
Phys. Lett. B \textbf{260} (1991), 186-192
doi:10.1016/0370-2693(91)90989-4.

\bibitem{Nielsen:1977sy}
N.~K.~Nielsen,
``The Energy Momentum Tensor in a Nonabelian Quark Gluon Theory,''
Nucl. Phys. B \textbf{120}, 212-220 (1977)
doi:10.1016/0550-3213(77)90040-2.

\bibitem{Adler:1976zt}
S.~L.~Adler, J.~C.~Collins and A.~Duncan,
``Energy-Momentum-Tensor Trace Anomaly in Spin 1/2 Quantum Electrodynamics,''
Phys. Rev. D \textbf{15}, 1712 (1977)
doi:10.1103/PhysRevD.15.1712.

\bibitem{Collins:1976yq}
J.~C.~Collins, A.~Duncan and S.~D.~Joglekar,
``Trace and Dilatation Anomalies in Gauge Theories,''
Phys. Rev. D \textbf{16}, 438-449 (1977)
doi:10.1103/PhysRevD.16.438.

\bibitem{Minkowski:1976en}
P.~Minkowski,
``On the Anomalous Divergence of the Dilatation Current in Gauge Theories,''
PRINT-76-0813 (BERN).


\bibitem{Tarrach:1981bi}
R.~Tarrach,
``The renormalization of FF,''
Nucl. Phys. B \textbf{196}, 45-61 (1982)
doi:10.1016/0550-3213(82)90301-7.

\bibitem{Espriu:1982bw}
D.~Espriu and R.~Tarrach,
``Renormalization of the Axial Anomaly Operators,''
Z. Phys. C \textbf{16}, 77 (1982)
doi:10.1007/BF01573750.

\bibitem{Freedman:1974gs}
D.~Z.~Freedman, I.~J.~Muzinich and E.~J.~Weinberg,
``On the Energy-Momentum Tensor in Gauge Field Theories,''
Annals Phys. \textbf{87}, 95 (1974)
doi:10.1016/0003-4916(74)90448-5.

\bibitem{Freedman:1974ze}
D.~Z.~Freedman and E.~J.~Weinberg,
``The Energy-Momentum Tensor in Scalar and Gauge Field Theories,''
Annals Phys. \textbf{87}, 354 (1974)
doi:10.1016/0003-4916(74)90040-2.

\bibitem{Hatta:2018sqd}
Y.~Hatta, A.~Rajan and K.~Tanaka,
``Quark and gluon contributions to the QCD trace anomaly,''
JHEP \textbf{12}, 008 (2018)
doi:10.1007/JHEP12(2018)008
[arXiv:1810.05116 [hep-ph]].

\bibitem{Furry:1951bef}
W.~H.~Furry,
``On Bound States and Scattering in Positron Theory,''
Phys. Rev. \textbf{81}, no.1, 115 (1951)
doi:10.1103/PhysRev.81.115.


\bibitem{Sapirstein:1990gn}
J.~R.~Sapirstein and D.~R.~Yennie,
``Theory of hydrogenic bound states,''
Adv. Ser. Direct. High Energy Phys. \textbf{7}, 560-672 (1990)
doi:10.1142/9789814503273\_0012.

\bibitem{Weinberg:1995mt}
S.~Weinberg,
``The Quantum theory of fields. Vol. 1: Foundations,''
Cambridge University Press, 2005,
ISBN 978-0-521-67053-1, 978-0-511-25204-4
doi:10.1017/CBO9781139644167.

\bibitem{blp1982}  V.~B.~Berestetskii, E.~M.~Lifshitz, and L.~P.~Pitaevskii, Quantum Electrodynamics: Volume 4 (Course of Theoretical Physics),  2nd Edition, Butterworth-Heinemann; 1982.

\bibitem{fock1930}  V.~A.~Fock, ''Bemerkung zum Virialsatz'', Z. Physik \textbf{63}, 855–858 (1930) doi.org/10.1007/BF01339281;  V. A. Fock, Selected  works, p.136, ed. L.D. Faddeev, L.A. Khalfin, I.V. Komarov, CRC Press, 2019,  ISBN-10: 0367394308.

\bibitem{shabaev1991} V.~M.~Shabaev, ''Generalizations of the virial relations for the Dirac
equation in a central field and their applications to
the Coulomb field'', J. Phys. B: At. Mol. Opt. Phys. \textbf{24}, 4479 (1991).

\bibitem{shabaev2002} V. M. Shabaev, ''Virial Relations for the Dirac Equation and
Their Applications to Calculations of Hydrogen-Like Atoms'' in: "Precision Physics of Simple Atomic Systems", edited by S.G. Karshenboim and V.B. Smirnov, (Berlin, Springer 2003) p. 97.  	arXiv:physics/0211087.

\bibitem{Yang:2014qna}
Y.~B.~Yang, Y.~Chen, T.~Draper, M.~Gong, K.~F.~Liu, Z.~Liu and J.~P.~Ma,
``Meson Mass Decomposition,''
PoS \textbf{LATTICE2014} (2014), 137
doi:10.22323/1.214.0137
[arXiv:1411.0927 [hep-lat].


\bibitem{Diakonov:1985eg}
D.~Diakonov and V.~Y.~Petrov,
``A Theory of Light Quarks in the Instanton Vacuum,''
Nucl. Phys. B \textbf{272}, 457-489 (1986)
doi:10.1016/0550-3213(86)90011-8.

\bibitem{akhber1981} A.~I.~Akhiezer and V.~B.~Berestetsky,  Quantum Electrodynamics, GFML, Nauka, Moscow 1981 (in Russian).

\bibitem{Mohr:1998grz}
P.~J.~Mohr, G.~Plunien and G.~Soff,
``QED corrections in heavy atoms,''
Phys. Rept. \textbf{293}, no.5-6, 227-369 (1998)
doi:10.1016/s0370-1573(97)00046-x.

\bibitem{Rose:1961} M.E. Rose, Relativistic Electron Theory, Wiley, New York, 1961.



\end{thebibliography}
\end{document}